\documentclass[11pt]{article}
\usepackage{amsmath}
\usepackage{comment}
\usepackage{amsfonts}
\usepackage{amsmath, amsfonts}
\usepackage{dsfont}
\usepackage{mathrsfs} 
\usepackage[active]{srcltx}
\usepackage{hyperref}
\usepackage[all]{xy}
\usepackage{bm}
\usepackage{dcolumn}
\usepackage[dvipsnames]{xcolor}
\usepackage{graphicx}
\usepackage{siunitx}
\usepackage{yfonts}
\usepackage{textcomp}
\usepackage{textgreek}
\usepackage{physics}
\usepackage[cal=boondox]{mathalfa}
\usepackage[bottom]{footmisc}
\usepackage{amssymb}
\usepackage{mathtools}
\usepackage{tikz}
\usepackage{bigints}

\usepackage[utf8]{inputenc}
\usepackage{array}
\usepackage{multirow}
\usepackage{tabu}

\makeatletter
\newcommand*{\rom}[1]{\expandafter\@slowromancap\romannumeral #1@}
\makeatother

\usepackage{fancyhdr} 
\usepackage[utf8]{inputenc}
\usepackage[english]{babel}

\DeclareMathAlphabet{\mathpzc}{OT1}{pzc}{m}{it}
\graphicspath{ {./images/} }

\hypersetup{
    colorlinks=true,   
    linkcolor=black,
    urlcolor=blue,
    citecolor=black
}
\usepackage{color}
\definecolor{DarkBlue}{rgb}{0.1,0.1,0.5}
\definecolor{Red}{rgb}{0.9,0.1,0.1}
\definecolor{Green}{rgb}{0.3,0.7,0.0}
\definecolor{green2}{rgb}{0.1,0.7,0.2}
\definecolor{blue2}{rgb}{0.0,0.6,0.7}
\definecolor{pink}{rgb}{1,0.0,1}
\definecolor{orange}{rgb}{0.9,0.0,0.1}
 
\usepackage[bottom]{footmisc}

\newtheorem{theo}{Theorem}

\newtheorem{prop}{Proposition}

\newtheorem{definition}{Definition}
\newtheorem{remark}{Remark}

\renewcommand{\d}{\mathrm{d}}
\newcommand{\derpar}[2]{\displaystyle\frac{\partial{#1}}{\partial{#2}}}

\newcommand{\Lag}{\mathscr{L}}

\newcommand{\vf}{\mathfrak{X}}
\newcommand{\df}{\Omega}

\newcommand{\Tan}{\mathrm{T}}
\newcommand{\inn}{{\mathop{i}\nolimits}}
\newcommand{\Lie}{\mathop{\mathrm{L}}\nolimits}

\def\qed{\ifvmode\removelastskip\fi
{\unskip\nobreak\hfil\penalty50\hbox{}\nobreak\hfil \hbox{\vrule
height1.2ex width1.2ex}\parfillskip=0pt \finalhyphendemerits=0
\par\smallskip}}

\newcommand{\bal}{\begin{align*}}
\newcommand{\eal}{\end{align*}}
\def\beq{\begin{equation}}
\def\eeq{\end{equation}}
\def\bea{\begin{eqnarray}}
\def\eea{\end{eqnarray}}
\def\beann{\begin{eqnarray*}}
\def\eeann{\end{eqnarray*}}
\def\ben{\begin{enumerate}}
\def\een{\end{enumerate}}
\def\bit{\begin{itemize}}
\def\eit{\end{itemize}}

\def\dst{\displaystyle}

\def\vf{\mathfrak X}
\def\df{{\mit\Omega}}
\def\Diff{{\rm Diff}}
\def\d{{\rm d}}

\def\Real{\mathbb{R}}

\def\Tan{{\rm T}}
\def\Lie{\mathop{\rm L}\nolimits}
\def\inn{\mathop{i}\nolimits}
\def\Cinfty{{\rm C}^\infty}

\title{\vskip -10mm
\sc More insights into symmetries in multisymplectic field theories
}
\author{\sffamily 
\sc 
$^a$ Arnoldo Guerra IV
\thanks{aguerrgu54@alumnes.ub.edu\,({\it ORCID}:\,0000-0001-8738-274X).} ,
$^b$ Narciso Rom\'an-Roy
\thanks{narciso.roman@upc.edu\,({\it ORCID}:\,0000-0003-3663-9861).} .
\\[1ex]
\normalsize\itshape\sffamily 
$^a $Departament de F\'isica Qu\`antica i Astrof\'isica,
Universitat de Barcelona, 
Barcelona, Spain.
\\[1ex]
\normalsize\itshape\sffamily 
$^b $Department of Mathematics,
Universitat Polit\`ecnica de Catalunya,
Barcelona, Spain.
}

\begin{document}
\markright{\rm A. Guerra IV, N. Rom\'an-Roy\/:
   \sl More insights into symmetries in multisymplectic field theories}
\maketitle

\leavevmode
\vadjust{\kern -15mm}

\begin{abstract}
\noindent
This work provides a general overview for the treatment of symmetries in classical field theories and (pre)multisymplectic geometry. 
The geometric characteristics of the relation between how symmetries are interpreted in theoretical physics and in the geometric formulation of these theories are clarified. 
Finally, a general discussion is given on the structure of symmetries in the presence of constraints appearing in singular field theories.
Symmetries of some typical theories in  theoretical physics are analyzed through the construction of the relevant multimomentum maps which are the conserved quantities (by Noether's theorem) on the (pre)multisymplectic phase spaces.
\end{abstract}


\noindent\textbf{MSC\,2020 codes:}
{\it Primary}: 53D42, 70S05, 83C05; {\it Secondary}: 35Q75, 35Q76, 53Z05, 70H50,  83C99.


\bigskip
\setcounter{tocdepth}{2}
{
\small
\def\addvspace#1{\vskip 1pt}
\parskip 0pt plus 0.1mm
\tableofcontents
}

\section{Introduction}

One of the main research topics in mathematical physics 
is the study of symmetries in dynamical (mechanical) systems and field theories. From a more mathematical point of view, the such studies of symmetries involves the analysis
of ordinary and partial differential equations;
this is because symmetries are associated with 
conserved quantities, or conservation laws, which give fundamental information about the physical system.
The foundational work on symmetries in physical systems is that of 
{\it Emmy Noether},
\cite{KS-2011}.
The general underlying concept of symmetry in a physical system
was first presented as the invariance of the equations of motion under a transformation on the phase space of the system.
In the more modern geometric formulation of classical mechanics and classical field theories, symmetries are usually characterized by demanding the
invariance of some underlying geometric structure 
from which the preservation of the equations of motions is included as a consequence.

The {\sl multisymplectic formulation} of classical field theories takes place on the 
the {\sl multivelocity} and {\sl multimomentum phase spaces} where the Lagrangian and De Donder--Weyl Hamiltonian formulations are developed. These phase spaces are
fiber bundles $\varrho\colon\mathcal{M}\to M$
over an orientable $m$-dimensional manifold $M$ (typically spacetime) where $m>1$. In particular, $\mathscr{M}$ is either a jet bundle in the Lagrangian formalism,
or a bundle of forms (or a quotient of them) in the Hamiltonian formalism. This formulation of classical field theories can be viewed as a geometric extension of the classical mechanics of non-autonomous systems for which $m=1$.
Furthermore, the phase spaces for field theories are endowed with a characteristic geometric structure:
a multisymplectic or premultisymplectic form $\Omega\in\df^{m+1}(\mathscr{M})$
(depending on the regularity of the theory).
In practice, the (pre)multisymplectic structures are constructed by starting from the Lagrangian function of the field theory under investigation
\cite{art:Aldaya_Azcarraga78_2,EMR-96,Gc-73,GS-73,HK-04,
book:Saunders89}.
Then, the field equations 
(the Euler-Lagrange equations or the the Hamilton- De Donder-Weyl equations)
are stated geometrically using the corresponding (pre)multisymplectic forms.

All symmetries, along with their associated conserved quantities and conservation laws,
have been studied extensively in the geometric framework,
both in the Lagrangian setting
\cite{art:Aldaya_Azcarraga78_2,De-77,art:deLeon_etal2004,
EMR-96,FS-2012,Gc-73,GMS-97,GS-73,Krupka,book:Saunders89}
and in the corresponding De Donder--Weyl Hamiltonian setting
\cite{art:deLeon_etal2004,EMR-99b,HK-04},
where the symmetries which preserve the (pre)multisymplectic forms are called {\sl Noether} or {\sl Cartan symmetries}. 
The geometric presentation of Noether symmetries in field theories
culminates with the statement of the geometric version of Noether’s theorem \cite{art:deLeon_etal2004,EMR-99b,Gc-73,art:GPR-2016,GR-3,GIMMSY,RWZ-2016}.
 We have paid special attention to the {\sl multimomentum maps} which are the fundamental conserved quantities associated with Noether symmetries.
These are the generalization of the {\sl momentum maps} of symplectic mechanics to the multisymplectic setting for field theory. 
(Multi)momentum maps are the relevant quantities for performing a {\sl symmetry reduction procedure}. 
However, the multisymplectic reduction of field theories is  currently, in general, an unsolved problem under research \cite{Bl-21,EMR-18,MaSw1,Sn-2004}.
The analogous geometric construction of symmetries in (first-order) non-autonomous mechanical systems for which $M=\Real$, the corresponding results for the Lagrangian and Hamiltonian formalisms can be found, for instance, in \cite{Ar,MS-98,SC-81}.

Singular field theories are invariant under a family of Noether symmetries called {\sl geometric gauge symmetries} which are linked to the degeneracy of the Lagrangian and are generated on the premultisymplecic phase spaces by vector fields which lie in the kernel of the premultisymplectic forms.
However, the term ``gauge symmetry" is used in various different contexts throughout the literature \cite{Bl-81}; such contexts will be discussed in section \ref{gauge}.
The so-called geometric gauge symmetries form {\sl gauge orbits} on the premultisymplectic phase space which define equivalence classes. Physical states which lie in the same gauge orbit differ from one another by a geometric gauge transformation and are said to be {\sl gauge equivalent}.

This paper has three main aims which are as follows. 
The first aim of this work is to review and broaden the discussion 
about the geometric structures and properties regarding 
the study of symmetries of first-order classical field theories 
and is mainly based on the previous papers \cite{art:GPR-2016,GR-3}.
The second main goal of this work is to clarify and enhance the relationship between the geometric treatment of classical field theories found in the mathematics literature and the equivalent approaches which are standard in theoretical physics. 
We believe this second goal is especially important as there seems to be several gaps in the dialogue between the communities of theoretical physics and differential geometry in this regard. In the theoretical physics literature it is standard to work on spaces of sections of some of the fiber bundles mentioned here while in the differential geometry literature the analysis of field theories is conducted on the fiber bundles themselves. 
For example, in theoretical physics it is standard to develop the Lagrangian formulation of field theories on the space of sections of the multivelocity phase space where
the Lagrangian function is taken to be a functional of the fields and its spacetime derivatives, while in the mathematics literature, typically, the Lagrangian function is taken to be a function on the multivelocity phase space itself as mentioned earlier.
The equivalence between these approaches will be made precise throughout the paper with sections \ref{liftME} and \ref{liftingvf} being of upmost importance to this aim.
Finally, the third aim of this paper is to detail some of the subtleties that arise in the analysis of symmetries when dealing with field theories that are singular in the De Donder--Weyl sense.  
The main goals of this paper mentioned above will be further pursued through the geometric analysis of specific field theories that are relevant in theoretical physics.

This paper is organized as follows.
Section \ref{syms} reviews the main definitions and results regarding the various types of 
symmetries along with the corresponding conserved quantities and  multimomentum maps
in the geometric framework of generic (pre)multisymplectic bundles.
Some new clarifying ideas regarding Noether and gauge symmetries are also provided.
Section \ref{lifting} is devoted to explaining how to lift diffeomorphisms and vector fields from the base space to the corresponding jet bundle;
this is important in order to characterize the most common types of symmetries in physics.
In Section \ref{syms2}, which contains the main contribution of the work,
the definitions and results of Section \ref{syms} 
are adapted to the Lagrangian and Hamiltonian descriptions of classical field theories. 
Aditionally, the so-called {\sl Lagrangian symmetries} arising in Lagrangian field theories are discussed.
The connection with the physical interpretation of all these ideas is detailed.
Finally, in Section \ref{examples}, the symmetries of some typical theories in theoretical physics are displayed.
Basic notions on multivector fields are given in appendix \ref{append}.

The classical field theories discussed here exhibit various kinds of symmetries that are typical in theoretical physics. 
The Nambu-Goto action for bosonic strings, which is invariant under the full group of diffeomorphisms (local point transformations) of the string worldsheet, is studied in the multisymplectic setting in Section \ref{Ex2}. 
The string action can be thought of as a matter action of a two dimensional field theory coupled to gravity; it is well known that, in this sense, the Nambu--Goto action is invariant only under spacetime isometries and not under the full group of spacetime diffeomorphisms as is General Relativity. 
This is typical for matter theories, including Yang--Mills which is presented subsequently. 
The symmetry of spacetime isometries, as well as the $SU(N)$ gauge symmetry, of Yang--Mills is discussed Section \ref{YangMills}. 
Chern--Simons theory with gauge group $SU(N)$
is the only theory discussed in this work (Section \ref{Ex3}) that is invariant under the full group of spacetime diffeomorphisms. 
Finally, in Sections \ref{electriccarroll} and \ref{Ex5}, we discuss the electric and magnetic Carrollian scalar field theories which are invariant only under Carroll transformations of spacetime; for more information on Carrollian field theories see, for example, \cite{CarrollScalar} and references therein.

Some comments about notation: 
$\vf(\mathscr{M})$ and $\df^k(\mathscr{M})$ 
denote the $\Cinfty(\mathscr{M})$-modules of vector fields and differential $k$-forms on a manifold $\mathscr{M}$ respectively,
$\Lie(X)\Omega$ donotes the Lie derivative of a differential form $\Omega$ with respect to a vector field $X$,
and $\inn(X)\Omega$ is the inner contraction between $X$ and $\Omega$.

All manifolds are taken to be real, second countable, and $\Cinfty$. All maps are also $\Cinfty$.  Sum over repeated indices (Einstein notation) is understood.


\section{Symmetries on (pre)multisymplectic fiber bundles}
\label{syms}

This section is devoted to the introduction of the main concepts and properties
regarding symmetries and conserved quantities (i.e. conservation laws) on (pre)multisymplectic manifolds.
This presentation is made in the generic geometric framework of (pre)multisymplectic fiber bundles which are the relevant structures on which classical field theories are constructed. The structures discussed in this section will later be particularized
to the specific contexts of the Lagrangian and Hamiltonian formalisms for classical field theories.
The fundamental ideas in this section are taken from \cite{art:GPR-2016}. 
For the proofs of the results and other details, see also \cite{EMR-99b,art:GPR-2016,GR-3}.

\subsection{(Pre)multisymplectic bundles}
\label{pmb}

Given a differentiable manifold $\mathscr{M}$,
a differential form $\Omega\in\df^m(\mathscr{M})$ is
{\sl $1$-nondegenerate} if, for every ${\rm p}\in\mathscr{M}$ and $Y\in\vf(\mathscr{M})$,
it follows that $\inn(Y)\Omega\vert_{\rm p}=0\ \Longleftrightarrow\ Y\vert_{\rm p}=0$.
Then, $\Omega\in\df^m(\mathscr{M})$ is said to be a {\sl\textbf{multisymplectic form}} 
if it is closed and $1$-nondegenerate and $(\mathscr{M},\Omega)$ is called a {\sl\textbf{multisymplectic manifold}}.
Alternatively, $\Omega$ is called a {\sl\textbf{premultisymplectic form}} if it is closed and $1$-degenerate 
and $(\mathscr{M},\Omega)$ is called a {\sl\textbf{premultisymplectic manifold}}.
If $\Omega$ is an exact form, then it is called an {\sl\textbf{exact (pre)multisymplectic form}} and the couple $(\mathscr{M},\Omega)$ is called a {\sl\textbf{(pre)multisymplectic system}}.

The geometric framework for Lagrangian and Hamiltonian field theories
consists of a fiber bundle $\varrho\colon \mathcal{M}\to M$
over an orientable manifold $M$ ($\dim\,M=m>1$, $\dim\,\mathscr{M}=N+m$).
The volume form on $M$ is denoted $\omega\in\df^m(M)$
(with $\omega$ also denoting $\varrho^*\omega$),
and $\mathscr{M}$ is endowed with an exact multisymplectic or premultisymplectic form 
(depending on the regularity of the theory) $\Omega=-\d\Theta\in\df^{m+1}(\mathscr{M})$ for some $\Theta\in\df^m(\mathscr{M})$.
Furthermore, the (pre)multisymplectic form satisfies the {\sl variational condition}
$\inn(Z_1)\inn(Z_2)\inn(Z_3)\Omega=0$, for all
$\varrho$-vertical vector fields $Z_1,Z_2,Z_3\in\vf^{V(\varrho)}(\mathscr{M})$,
which allows for the field equations to be obtained from a variational principle.

The solution to the variational problem on $\mathscr{M}$ are sections
$\bm{\psi}\colon M\to\mathscr{M}$ of the projection $\varrho$ which solve the field equations given by
$$
\bm{\psi}^*\inn(X)\Omega= 0 \quad , \quad \text{for every }X\in\vf(\mathscr{M}) \ .
$$
Equivalently, the solutions to the variational problem are integral sections of a $m$-multivector field
(see appendix \ref{append})
contained in a class of integrable, $\varrho$-transverse (i.e. ${\bf X}\in\ker^m\,\Omega$), $m$-multivector fields
$\left\{ {\bf X} \right\} \subset \vf^m(\mathscr{M})$
which satisfy the field equations which are now written as
\beq
\inn({\bf X})\Omega=0 \ ,
\label{fundeqs1}
\eeq
where the $\varrho$-transversality condition can be set, without loss of generality, as 
\beq
\inn({\bf X})\omega=1\ .
\label{fundeqs0}
\eeq
The couple $(\mathscr{M},\Omega)$ is usually called a {\sl (pre)multisymplectic system}.

In general, solvable premultisymplectic systems admit stable solutions to the field equations on some submanifold ${\cal S}\subseteq\mathscr{M}$
which is obtained by applying a {\sl constraint algorithm} (see Section \ref{proj} and \cite{LMMMR-2005} for more details);
the submanifold ${\cal S}$ is called the {\sl final constraint submanifold} 
and the physical states of the field theory under investigation are the sections of the projection $\varrho$ whose images are on ${\cal S}$.

\subsection{Conserved quantities and symmetries}

In geometric mechanics, conserved quantities are functions on the phase space which are invariant under the advance along the dynamical trajectories given by the integral curves of some dynamical vector fields.
The invariance condition is stated by demanding the vanishing of the Lie derivative of the functions representing the conserved quantity 
with respect to the dynamical vector field solution to the equations of motion.
For field theories, there is an analogous geometrical way 
to introduce this concept using multivector fields 
to represent the solutions to the field equations.

Let $(\mathscr{M},\Omega)$ be a (pre)multisymplectic system.

\begin{definition}
A \textbf{conserved quantity} of the (pre)multisymplectic system
$(\mathscr{M},\Omega)$ is a form $\alpha\in\df^{m-1}(\mathscr{M})$ which satisfies
$\Lie({\bf X})\alpha=(-1)^{m+1}\inn({\bf X})\d\alpha=0$ 
for every locally decomposable and $\varrho$-transverse multivector field ${\bf X}\in\ker^m\Omega$ 
(i.e., which satisfies equations \eqref{fundeqs1} and \eqref{fundeqs0}).
 \end{definition}
Conserved quantities are characterized by the following property:
if $\alpha\in\df^{m-1}(\mathscr{M})$ is a conserved quantity and ${\bf X}\in\ker^m\Omega$
is a $\varrho$-transverse integrable multivector field,
then $\alpha$ is closed on the integral submanifolds of ${\bf X}$; that is,
if $j_S\colon S\hookrightarrow \mathscr{M}$
 is an integral submanifold of ${\bf X}$, then $\d j_S^*\alpha=0$.

\begin{remark}
\label{current}
{\rm Conserved quantities in field theories appear as {\sl conservation laws} or {\sl conserved currents} as,
for every $\alpha\in\df^{m-1}(\mathscr{M})$ and ${\bf X}\in\vf^m(\mathscr{M})$,
if ${\bm\psi}\colon M\to \mathscr{M}$ is an  integral section of ${\bf X}$ such that ${\bm\psi}^*\alpha\in\df^{m-1}(M)$,
then there is a unique vector field $X_{{\bm\psi}^*\alpha}\in\vf(M)$ such that
$\inn(X_{{\bm\psi}^*\alpha})\eta={\bm\psi}^*\alpha$.
It follows that the {\sl divergence} of $X_{{\bm\psi}^*\alpha}$ is the function ${\rm div}X_{{\bm\psi}^*\alpha}\in\Cinfty(M)$
defined as $\Lie(X_{{\bm\psi}^*\alpha})\omega= ({\rm div}X_{{\bm\psi}^*\alpha})\,\omega$ which
 satisfies 
$({\rm div}X_{{\bm\psi}^*\alpha})\,\eta=\d{{\bm\psi}^*\alpha}$.
Therefore, as a consequence of the above property,
$\alpha$ is a conserved quantity if,  and only if, ${\rm div}X_{{\bm\psi}^*\alpha}=0$
and hence {\sl Stokes theorem} assures that, in every bounded domain $U\subset M$,
$$
\int_{\partial U}{{\bm\psi}^*\alpha}=\int_U ({\rm div}X_{{\bm\psi}^*\alpha})\,\eta=\int_U\d{{\bm\psi}^*\alpha}=0 \ .
$$
The form ${\bm\psi}^*\alpha$ is called the {\sl \textbf{current}} associated with the conserved quantity $\alpha$.
This result associates 
a {\sl \textbf{conservation law}} on $M$ to every conserved quantity on $\mathscr{M}$.
} \end{remark}

The field equations are partial differential equations ({\sc pde}'s) and {\sl symmetries} are (local) diffeomorphisms that transform solutions to the field equations into equivalent solutions.
In geometric mechanics, this property is stated using the vector fields which generate the symmetries 
(by their flows) and the dynamical vector fields which solve the equations of motion.
For field theories, there is a similar geometric procedure 
using multivector fields 
which are solutions to the field equations 
as follows:

 \begin{definition}
 \label{defsymm}
 \ \
\ben
\item
A \textbf{symmetry} of the (pre)multisymplectic system $(\mathscr{M},\Omega)$
is a diffeomorphism $\Phi\colon \mathscr{M}\to \mathscr{M}$ 
such that $\Phi_*(\ker^m\Omega)\subset\ker^m\Omega$.
\item
 An \textbf{infinitesimal symmetry} of the (pre)multisymplectic system $(\mathscr{M},\Omega)$
is a vector field \, $Y\in\vf(\mathscr{M})$ whose local flows are 
local symmetries or, equivalently, \,
$[Y,\ker^m\Omega]\subset\ker^m\Omega$.
\een
 \end{definition}

\begin{remark}{\rm
If $(\mathscr{M},\Omega)$ is a premultisymplectic system with final constraint submanifold
$\jmath_{{\cal S}}\colon{\cal S}\hookrightarrow \mathscr{M}$, then symmetries must be diffeomorphisms
$\Phi\in{\rm Diff}(\mathscr{M})$ such that $\Phi({\cal S})={\cal S}$ 
and infinitesimal symmetries are vector fields $Y\in\vf (\mathscr{M})$
tangent to ${\cal S}$ which satisfy the conditions in Definition \ref{defsymm} at least on ${\cal S}$.
The set of vector fields on $\mathscr{M}$ which are tangent to ${\cal S}$ will be denoted here as 
$$
\underline{\vf({\cal S})}:=\{ Y\in\vf(\mathscr{M})\ \vert\ 
\exists Y_{\cal S}\in\vf({\cal S})\ \ \vert \ \ \jmath_{{\cal S}*}Y_{\cal S}=Y\vert_{\cal S}\} \ .
$$
The elements of $\underline{\vf({\cal S})}$ are the vector fields $Y\in\vf(\mathscr{M})$
which, for some $Y_{\cal S}\in\vf({\cal S})$, make the following diagram commutative:
 \[
 \begin{array}{cccc}
\begin{picture}(90,52)(0,0)
 \put(5,0){\mbox{${\cal S}$}}
 \put(5,42){\mbox{$\mathscr{M}$}}
 \put(10,13){\vector(0,1){25}}
 \put(-5,22){\mbox{$\jmath_{\cal S}$}}
 \put(30,45){\vector(1,0){55}}
 \put(30,4){\vector(1,0){55}}
 \put(48,12){\mbox{$Y_{\cal S}$}}
 \put(50,33){\mbox{$Y$}}
 \end{picture}
&
\begin{picture}(15,52)(0,0)
 \put(0,0){\mbox{$\Tan{\cal S}$}}
 \put(0,41){\mbox{$\Tan\mathscr{M}$}}
 \put(10,13){\vector(0,1){25}}
\put(15,22){\mbox{$\Tan\jmath_{\cal S}$}}
\end{picture}
 \end{array}
 \]
}	
\end{remark}

As we have said, symmetries transform solutions to the field equations onto solutions, since:

\begin{theo}
\label{gsymsol}
If $\Phi\in{\rm Diff}(\mathscr{M})$ is a symmetry that restricts to a diffeormorphism $\varphi\colon M\to M$;
then, for every $\varrho$-transverse integrable multivector field ${\bf X}$,
the map $\Phi$ transforms integral sections of ${\bf X}$
into integral sections of \,$\Phi_*{\bf X}$ 
and hence $\Phi_*{\bf X}\in\ker^m\Omega$
which is also a $\varrho$-transverse integrable multivector field.
Consequently,
if $Y\in\vf (\mathscr{M})$ is a $\bar\pi^1$-projectable infinitesimal symmetry
and $F_t$ is the local flow of $Y$, then
$F_t$ transforms integral sections of every
$\varrho$-transverse integrable multivector field ${\bf X}\in\ker^m\Omega$ 
into integral sections of $F_{t*}{\bf X}$.
\end{theo}

If $Y_1,Y_2\in\vf (\mathscr{M})$ are infinitesimal symmetries, then the Lie bracket $[Y_1,Y_2]$ is also an infinitesimal symmetry.
Furthermore, if $\Phi\in{\rm Diff}(\mathscr{M})$ is a symmetry and $\alpha\in\df^{m-1}(\mathscr{M})$
is a conserved quantity,  then $\Phi^*\alpha$ is also a conserved quantity.
Consequently,
if $Y\in\vf (\mathscr{M})$ is an infinitesimal symmetry,
then $\Lie(Y)\alpha$ is also a conserved quantity.

\begin{remark}
\label{fiberpreserve}
{\rm Transformations of the base space $M$  
(usually spacetime transformations in field theory)
which preserve the (pre)multisymplectic structure of the phase spaces are called {\sl \textbf{spacetime symmetries}}.
Therefore, in the bundle $\varrho\colon \mathscr{M}\to M$, the corresponding diffeomorphisms $\Phi\colon \mathscr{M}\to \mathscr{M}$
must be fiber preserving and thereby restrict to diffeomorphisms $\Phi_M\colon M\to M$
which satisfy $\Phi_M\circ\varrho=\varrho\circ\Phi$.
For infinitesimal symmetries, this means that
the vector fields $Y\in\vf(\mathscr{M})$ must be
$\varrho$-projectable, hence,
there exist $Y_M\in\vf(M)$ such that
$\varrho_*Y=Y_M$.
} \end{remark}

Symmetries are generated by the {\sl action} of a Lie group $G$ on $\mathscr{M}$:
$$
\begin{array}{ccccc}
\Phi&\colon&G&\to&\Diff(\mathscr{M})
\\
& &g&\mapsto&\Phi_g
\end{array}
$$
where $\Phi_g:\mathscr{M}\rightarrow \mathscr{M}$ are the local diffeomorphisms induced by the group action. 
If $\mathfrak{g}$ is the Lie algebra of $G$, it follows that every $\xi\in\mathfrak{g}$ 
induces a vector field $\textbf{X}_\xi\in\vf(\mathscr{M})$ which generates the diffeomorphisms produced by the group action of $G$ on $\mathscr{M}$. 
This is achieved by constructing the exponential map for some $g\in G$:
$$
\begin{array}{ccccc}
exp&\colon&\textgoth{g}&\to&G
\\
& &\xi&\mapsto&g=exp(\lambda\xi)
\end{array}
$$
where $\lambda\in\Real$ is a parameter. 
Then, the induced vector field $X_\xi$ on $\mathscr{M}$ is written in terms of the exponential map as
\begin{equation*}
\left. X_{\xi}=\frac{\text{d}}{\text{d}\lambda}\right\vert_{\lambda=0}\Phi^*_{exp(\lambda\xi)} \in\vf(\mathscr{M})\ .
\end{equation*}
These vector fields are the {\sl infinitesimal generators} of the symmetries, that is, the infinitesimal symmetries.

\subsection{Noether symmetries}

In order to obtain conserved quantities (or conservation laws) associated with symmetries,
it is necessary to impose some additional conditions.
In general, these conditions are related to the geometric structures underlying the field equations
(see, for instance, \cite{RR-2020,SC-81} and the references therein, for a complete study in Lagrangian and Hamiltonian mechanics).
The most relevant kinds of symmetries are the following:

 \begin{definition}
 \label{Cartansym}
 \ \
\ben
\item
A \textbf{Noether} or \textbf{Cartan symmetry}
of the (pre)multisymplectic system $(\mathscr{M},\Omega)$
is a diffeomorphism $\Phi\colon \mathscr{M}\to \mathscr{M}$
such that, $\Phi^*\Omega=\Omega$.
In the particular case where $\Phi^*\Theta=\Theta$, then $\Phi$ is called an
\textbf{exact Noether} or \textbf{exact Cartan symmetry}.
\item
An \textbf{infinitesimal Noether} or \textbf{ Cartan symmetry} of the (pre)multisymplectic system $(\mathscr{M},\Omega)$
is a vector field $Y\in\vf (\mathscr{M})$ for which $\Lie(Y)\Omega=0$.
In the particular case where $\Lie(Y)\Theta=0$, $Y$ is called an
\textbf{infinitesimal exact Noether} or \textbf{infinitesimal exact Cartan symmetry}.
\een
 \end{definition}
 
If $Y_1,Y_2\in\vf (\mathscr{M})$
are infinitesimal Noether symmetries, then so is the Lie bracket $[Y_1,Y_2]$.
Furthermore, if $Y\in\vf(\mathscr{M})$ is an infinitesimal Noether symmetry
and ${\bf X}\in\ker^m\Omega$, bearing in mind \eqref{liebrac} we have that
$$
 \inn([Y,{\bf X}])\Omega= \Lie(Y)\inn({\bf X})\Omega-\inn({\bf X})\Lie(Y)\Omega=0 \ \Longleftrightarrow \ [Y,{\bf X}]\subset\ker^m\Omega \ ,
$$
and hence infinitesimal Noether symmetries are infinitesimal symmetries (and the same holds for Noether symmetries).

\begin{remark}
\label{consquant}
{\rm
As $\Lie(Y)\Omega_{\Lag}=\d\inn(Y)\Omega_{\Lag}$,
the condition $\Lie(Y)\Omega=0$
 is equivalent to demanding that $\inn(Y)\Omega$
 is a closed $m$-form on $\mathscr{M}$. It follows that
$\inn(Y)\Omega=-\d\alpha_Y$ is an open set $U\subset \mathscr{M}$
for some $\alpha_Y\in\df^{m-1}(U)$
(where the minus sign is due to physical conventions).
Thus, an infinitesimal Noether symmetry is a {\sl locally Hamiltonian vector field} for the (pre)multisymplectic form 
$\Omega$ and $\alpha_Y$ is the corresponding {\sl local Hamiltonian form}.}
\end{remark}

Therefore, {\sl \textbf{Noether's theorem}} is stated as follows:

 \begin{theo}
 {\rm (Noether):}
Let $Y\in\vf (\mathscr{M})$ be an infinitesimal Noether symmetry
with $\inn(Y)\Omega=-\d\alpha_Y$
in an open set $U\subset \mathscr{M}$. Then,
for every locally decomposable and $\varrho$-transverse (integrable) multivector field ${\bf X}\in\ker^m(\Omega$, it follows that
 $$
 \Lie({\bf X})\alpha_Y=0 \ ;
 $$
 that is, every Hamiltonian $(m-1)$-form
 $\alpha_Y$ associated with $Y$ is a conserved quantity.
For every integral section ${\bm\psi}$ of ${\bf X}$,
 the form ${\bm\psi}^*\alpha_Y$ is usually called a {\sl \textbf{Noether current}}.
 \label{Nth}
 \end{theo}

Observe that the form $\Lie(Y)\Theta$ is closed since
\beq
 \Lie(Y)\Theta=\d\inn(Y)\Theta+\inn(Y)\d\Theta=
\d\inn(Y)\Theta-\inn(Y)\Omega=
\d (\inn(Y)\Theta-\alpha_Y)\equiv\d\zeta_Y
 \quad \mbox{\rm (in $U$)} \ .
 \label{closedTheta}
\eeq
In particular, if $Y$ is an exact infinitesimal Noether symmetry, 
then $\zeta_Y$ is closed and $\alpha_Y=-\inn(Y)\Theta$.

\begin{remark}
\label{fiberpreserve2}
{\rm As it was pointed out in Remark \ref{fiberpreserve},
we will mainly be interested in Noether symmetries which are
are fiber-preserving diffeomorphisms for the bundle $\varrho\colon\mathscr{M}\to M$
and in infinitesimal Noether symmetries which are
$\varrho$-projectable vector fields.

Furthermore, if $(\mathscr{M},\Omega)$ is a premultisymplectic system with final constraint submanifold
${\cal S}\hookrightarrow \mathscr{M}$, then Noether symmetries 
must be diffeomorphisms
$\Phi\in{\rm Diff}(\mathscr{M})$ such that $\Phi({\cal S})={\cal S}$ 
and infinitesimal Noether symmetries must be vector fields $Y\in\vf (\mathscr{M})$
tangent to ${\cal S}$ which satisfy the conditions of Definition \ref{Cartansym}, at least on ${\cal S}$.
}
\end{remark}

\subsection{Gauge symmetries}
\label{gauge}

The term ``gauge" is used in various different contexts throughout the literature. 
In physics, it is typically used to refer to field transformations which are spacetime dependent.
This section discusses some of the different meanings associated with the term ``gauge" within the context of classical field theory. 
In particular, the description of a certain type of {\sl gauge symmetries} that are the most relevant from both a geometric and physical point of view will be given.
See also \cite{Ga-2022,GR-3} for additional insights.

Let $(\mathscr{M},\Omega)$ be a (pre)multisymplectic system and recall that Noether symmetries preserve the (pre)multisymplectic form $\Omega$.
Only fiber-preserving symmetries will be investigated, bearing in mind Remarks \ref{fiberpreserve} and \ref{fiberpreserve2} which characterize symmetries in the context of the final constraint submanifold ${\cal S}\subset\mathscr{M}$ of the premultisymplectic system under investigation.
Furthermore, the particular situation where such symmetries restrict to the identity on $M$ will be the kind which will be generically referred to here as {\sl \textbf{gauge symmetries}}.
Then, {\sl \textbf{infinitesimal gauge symmetries}}
are vector fields $Y\in\vf(\mathscr{M})$ whose local flows
are local gauge symmetries and, consequently,
are $\varrho$-vertical vector fields.
In this respect, it should be noted that, the set of $\varrho$-vertical vector fields
are $\vf^{V(\varrho)}(\mathscr{M})=\ker\,\omega$.

Furthermore, infinitesimal Noether symmetries are vector fields $Y\in\vf(\mathscr{M})$
characterized by the the property that $\d\inn(Y)\Omega=0$.
When $\ker\,\Omega\not=\{ 0\}$
then $(\mathscr{M},\Omega)$ is a premultisymplectic system. Then, any nonzero vector field $Y\in\ker\,\Omega$ (i.e., $\inn(Y)\Omega=0$) is
a particular type of infinitesimal symmetry that is related to the degeneracy of
the premultisymplectic form $\Omega$.
Such vector field will be referred to as {\sl\textbf{geometric (infinitesimal) gauge symmetries}}.

Moreover, when $(\mathscr{M},\Omega)$ is a premultisymplectic system, the field equations, in general, have consistent solutions 
on a final constraint submanifold  ${\cal S}\hookrightarrow \mathscr{M}$
(as stated at the end of Section \ref{pmb}) 
which is the manifold where sections representing the physical states take their image. 
As before, the physically relevant gauge transformations are those which transform physical states into equivalent physical states and hence
are automorphisms of the submanifold ${\cal S}$ which transform sections
of the projection $\varrho$ into themselves.
Consequently, these gauge transformations which act along the fibers of the projection $\varrho$ are necessarily generated by $\varrho$-vertical vector fields.

The concepts above are summarized as follows:

\begin{definition}
A \textbf{geometric infinitesimal gauge symmetry}, or simply a \textbf{gauge vector field}, 
of a (pre)multisymplectic system $(\mathscr{M},\Omega)$ is a vector field $Y\in\vf (\mathscr{M})$ such that:
\ben
\item
$Y\in\ker\,\Omega$.
\item
It is a $\varrho$-vertical vector field,  $Y\in\vf^{V(\varrho)}(\mathscr{M})$.
\item
It is tangent to ${\cal S}$,
$Y\in\underline{\vf({\cal S})}$.
\een
The set of gauge vector fields is denoted ${\cal G}$.
\end{definition}

The local diffeomorphisms generated by the flow of gauge vector fields 
are called {\sl \textbf{geometric gauge transformations}}.
Physical states (i.e., sections which are stable solutions to the field equations) 
related to one another by geometric gauge transformations are called {\sl \textbf{geometric gauge equivalent states}}
and they are physically equivalent
(in the sense that they are physically indistinguishable).

\begin{remark} {\rm
In the case where $\jmath_{\cal S}\colon{\cal S}\hookrightarrow\mathscr{M}$
is strictly a submanifold of $\mathscr{M}$,
besides those of $\ker\,\Omega$, there are more vector fields related to the non-regularity of the premultiplectic form $\Omega$.
Such vector fields belong to the set
$$
\underline{\ker\Omega_{\cal S}}:=\{ Z\in\underline{\vf({\cal S})}\ \vert\ 
\exists Z_{\cal S}\in\ker\Omega_{\cal S}\ \ \vert \ \ \jmath_{{\cal S}*}Z_{\cal S}=Z\vert_{\cal S}\} \ ,
$$
where $\Omega_{\cal S}=\jmath_S^*\Omega$ which, in general, is also a premultisymplectic form.
The vector fields in $\underline{\ker\Omega_{\cal S}}$ are those which are tangent to ${\cal S}$ and can be identified on the points of ${\cal S}$
with the elements of $\ker\Omega_{\cal S}$.
Since $\ker\,\Omega\cap\underline{\vf({\cal S})}\subseteq\underline{\ker\Omega_{\cal S}}$, then for every $Z\in\ker\,\Omega\cap\underline{\vf({\cal S})}$,
it follows that
$$
\inn(Z)\Omega=0 \ \Longrightarrow\ 
0=\jmath_{\cal S}^*\inn(Z)\Omega=\inn(Z_{\cal S})\Omega_{\cal S} 
 \ \Longleftrightarrow\  Z_S\in \ker\,\Omega_{\cal S}\ ,
$$
and, as a consequence,
$$
{\cal G}\equiv\ker\,\Omega\cap\underline{\vf({\cal S})}\cap\vf^{V(\varrho)}(\mathscr{M})\subseteq\underline{\ker\Omega_{\cal S}}\cap\vf^{V(\varrho)}(\mathscr{M})\equiv\widehat{\cal G}\ .
$$
In the case of singular (autonomous) dynamical systems in mechanics, 
it is proved \cite{BK-86,GN-79}
that $\widehat{\cal G}$ is the complete set of gauge vector fields.
Then, the elements of ${\cal G}$ are called {\sl primary gauge vector fields} and those in $\widehat{\cal G}-{\cal G}$
are the {\sl secondary gauge vector fields}.
The analogous concept in classical field theories regarding the classification of $\widehat{\cal G}-{\cal G}$ 
as gauge vector fields
(i.e., if they relate geometric gauge equivalent states through their flows) is still an open question.
} \end{remark}

As usual, if $Z_1,Z_2\in{\cal G}$, then $[Z_1,Z_2]\in{\cal G}$ so ${\cal G}$ generates an involutive distribution on $\mathscr{M}$
and the quotient set $\widetilde{\cal S}={\cal S}/{\cal G}$,
assumed to be a differentiable manifold,
plays the relevant role in the so-called {\sl geometric gauge reduction procedure} \cite{BK-86,GN-79}
which consists of taking this reduced manifold as the set of ``real physical" states
and then removes the unphysical redundancy represented by the geometric gauge equivalent states.
Introducing the canonical projection
$\tilde\tau_{\cal S}\colon{\cal S}\to\widetilde{\cal S}$,
gauge equivalent states are sections of this projection.
An alternative way to remove the ``geometric gauge redundancy'' is by performing a {\sl gauge fixing} procedure
which consists of fixing sections of this projection.
A summarized description of these two procedures may be found in \cite{GR-3}. 
The most general method for gauge fixing in field theory involves the BRST-BV method (see, for example, \cite{Gomis-BRST2,Gomis-BRST1}).

As a last comment, it should be noted that
sometimes in physics a broader definition of (infinitesimal) gauge symmetry is handled which pertains to 
vector fields $Y\in\vf (\mathscr{M})$ such that $\d\inn(Y)\Omega=0$, which, in the terminology used in this work, 
are the so-called infinitesimal Noether symmetries.

\subsection{Multimomentum map}

Let $G$ be a {\sl group of Noether symmetries} for a (pre)multisymplectic system $(\mathscr{M},\Omega)$; that is,
$\Phi_g^*\Omega=\Omega$, for every $g\in G$, and
$\Lie(X_{\xi})\Omega= 0$, for every $\xi\in\mathfrak{g}$.
Therefore, as $\d\inn(X_{\xi})\Omega=0$,
there exists a form ${\rm J}_\xi\in\df^{m-1}(U)$ on an open neighbourhood $U\subset \mathscr{M}$ of every point in $\mathscr{M}$ such that, 
\begin{equation}
\label{equation:dJ}
\inn(X_\xi)\Omega=-\text{d}{\rm J}_\xi\ .
\end{equation}
This form ${\rm J}_\xi$ is determined up to some exact form $\text{d}\beta_{\xi}$ with $\beta_{\xi}\in\df^{m-2}(U)$
and coincides with the conserved quantity $\alpha_Y$ introduced in Remark \ref{consquant} and Noether's theorem \ref{Nth}. 
Furthermore,
\begin{equation}
\label{equation:dJcartan}
\text{d}{\rm J}_{\xi}=-\inn(X_{\xi})\Omega_\Lag=\inn(X_{\xi})\text{d}\Theta=\Lie(X_{\xi})\Theta-\text{d}\inn(X_{\xi})\Theta \ ,
\end{equation}
and, as $X_{\xi}$ is an infinitesimal Noether symmetry,
then \eqref{closedTheta} holds for $X_{\xi}$ and
$\Lie(X_{\xi})\Theta=\text{d}\zeta_{\xi}$, for some $\zeta_{\xi} \in \df^{m-1}(F)$. 
Now, by (\ref{equation:dJcartan}) it follows that 
$$
{\rm J}_{\xi}=-\inn(X_{\xi})\Theta+\zeta_{\xi}+\text{d}\beta_{\xi} \ .
$$
The $(m-1)$-form ${\rm J}_{\xi}$ can be taken to be linear in $\xi$ by taking the generators $\xi\in\textgoth{g}$ to be infinitesimal.
Therefore, for every point ${\rm p}\in \mathscr{M}$, define the linear map
$$
\begin{array}{ccccc}
\left.{\rm J}\right\vert_{\rm p}&\colon&\textgoth{g}^*&\to&\df^{m-1}(\Tan_{\rm p}\mathscr{M})
\\
& &\xi&\mapsto&{\rm J}\vert_{\rm p}(\xi):={\rm j}_{\xi}({\rm p})\ .
\end{array} 
$$
Then:

\begin{definition}
The map 
$$
\begin{array}{ccccc}
{\rm J}&\colon& \mathscr{M}&\to&\textgoth{g}^*\otimes\df^{m-1}(F)
\\
& & {\rm p}&\mapsto & {\rm J}({\rm p}):={\rm J}\vert_{\rm p}
\end{array}
$$
is called the \textbf{multimomentum map}
associated with the symmetry group $G$.
\end{definition}

The terminology ``multimomentum map'' is also used to refer the $(m-1)$-form ${\rm J}_\xi$ arising in \eqref{equation:dJ}. 
Furthermore, it is usual to specify the multimomentum map by using the natural pairing between $\textgoth{g}^*$ and $\textgoth{g}$ as 
${\rm J}_{\xi}({\rm p})=\left<{\rm J}({\rm p}),\xi\right>$.

Thus, Noether symmetries on $\mathscr{M}$ produce multimomentum maps on $\mathscr{M}$
and the so-called {\sl \textbf{Noether current}} ${\rm j}=j^\mu\text{d}^{m-1}x_\mu\in \df^{m-1}(\mathscr{M})$ of a symmetry 
is obtained as ${\rm j}={\bm\psi}^*{\rm J}_\xi$ (see also Remark \ref{current}).

\section{Lifting transformations from the base space of a jet bundle}
\label{lifting}

In classical field theories,
some symmetries are associated with diffeomorphisms on the base manifold $M$
(i.e., spacetime transformations in most cases)
and, in the case of first-order Lagrangian field theories where $\mathscr{M}$ is a first-order jet bundle, $J^1\pi\to E\to M$
(see Section \ref{S2}),
it is common to obtain the law of transformation in $J^1\pi$ from 
those diffeomorphisms on $M$.

This section discusses how to induce jet bundle transformations from diffeomorphisms on the base space. See \cite{EMR-96,book:Saunders89} for other details.

\subsection{First-order jet bundles}
\label{jet}

Begin by letting $\pi\colon E \longrightarrow M$ be a fiber bundle over an orientable $m$-dimensional manifold $M$ with $\dim E = m + n$ and $m>1$.
The {\sl $1$st-order jet bundle} $J^1\pi$ of the projection $\pi$ is the manifold of the $1$-jets (equivalence classes) of local sections of $\pi$, denoted as $\phi\in\Gamma(\pi)$.
Given a point $x\in M$, the points in $J^1\pi$ are denoted by $\bar y\equiv j^1_x\phi$
where $\phi\in\Gamma(\pi)$ is a representative of the equivalence class.
The resulting natural projections are
$$
\begin{array}{rcl}
\pi^1 \colon J^1\pi & \longrightarrow & E \\
j^1_x\phi & \longmapsto & \phi(x)
\end{array}
\quad \ \quad
\begin{array}{rcl}
\bar{\pi}^1 \colon J^1\pi & \longrightarrow & M \\
j^1_x\phi & \longmapsto & x
\end{array}
$$
where $\bar{\pi}^1=\pi\circ\pi^1$.
The volume form on $M$ as well as its pull-back $\bar\pi^{1*}\omega$
are denoted by $\omega$.
Local coordinates on $J^1\pi$ are denoted
$(x^\mu,y^i,y_\mu^i)$, with $0 \leqslant\mu \leqslant m-1$, $1 \leqslant i \leqslant n$;
and $(x^\mu,y^i)$ are local coordinates on $E$ adapted to the bundle structure so that
$\omega=\d x^0\wedge\ldots\wedge\d x^{m-1}\equiv\d^mx$.

The {\sl canonical lift} of a section $\phi\in\Gamma(\pi)$ to $J^1\pi$ is denoted $j^1\phi\in\Gamma(\bar{\pi}^1)$.
A section $\psi\in\Gamma(\bar{\pi}^1)$ is said to be {\sl \textbf{holonomic}}
if $\psi$ is the canonical lift of a section $\phi=\pi^1\circ\psi\in\Gamma(\pi)$, and hence, $\psi=j^1(\pi^1\circ\psi)$.

\subsection{Lifting transformations from $M$ to $E$}
\label{liftME}

Consider infinitesimal diffeomorphisms $\Phi_M\colon M\rightarrow M$ produced by the coordinate transformation $x'^\mu=x^\mu+\xi^\mu(x)$.
These transformations are generated by vector fields $\displaystyle Z=-\xi^\mu(x)\frac{\partial}{\partial x^\mu}\in\vf(M)$ 
and the $\pi$-projectable vector fields generating the corresponding transformations on the configuration manifold $E$, 
which are written in full generality as $(x^\mu,y^A)\rightarrow (x^\mu+\xi^\mu(x),y^i+\xi^i(x,y))$ with $\xi^i(x,y)\in\Cinfty(E)$, are given as
\begin{equation}
\label{diffsEvector}
Y\equiv Z_E=-\xi^\mu(x)\frac{\partial}{\partial x^\mu}-\xi^i(x,y)\frac{\partial}{\partial y^i}\in\vf(E)\ .
\end{equation}
The fields $y^i(x)$ which are given as components of the sections $\phi(x)=(x^\mu,y^i(x))$ transform as the Lie derivatives of the fields
with respect to the vector field $Z$ on $M$ (denoted by $\delta y^i(x)$) as
\begin{equation}
\label{liederivative}
\delta y^i(x)\equiv\Lie(Z) y^i(x)=y'^i(x)-y^i(x)=-\xi^\mu(x)\derpar{y^i}{x^\mu}(x) +\widetilde{\xi}^i(x)\ .
\end{equation}
These field transformations are sometimes referred to as the {\sl \textbf{local variation}} of the fields.
The functions $\widetilde{\xi}^i(x)$ are not related to the components $\xi^i(x,y)$ in \eqref{diffsEvector}, {\it a priori}.
The term 
$-\xi^\mu(x)\derpar{y^i}{x^\mu}$ in (\ref{liederivative}) is called the {\sl \textbf{transport term}} and $\widetilde{\xi}^i(x)$ is called the {\sl \textbf{global variation}} of the fields which is given by
\begin{equation*}
\widetilde{\xi}^i(x)= y'^i(x')-y^i(x)\ .
\end{equation*}

A geometrical interpretation of the local field variations is given as follows
\cite{GIMMSY,GP-2002,KSM-2011,Mi-2008}:

\begin{definition}
Consider a section $\phi\colon M\to E$ and let $Z_E\in\vf(E)$ be a $\pi$-projectable vector field which projects to $Z\in\vf(M)$ (and hence has the local expression \eqref{diffsEvector}).
The \textbf{generalized Lie derivative of the section $\phi$ by $Z$} is the map
 $\mathbb{L}(Z)\phi\colon M\to\Tan E$
  defined as
\begin{equation}
\label{LieDerivSections1}
\mathbb{L}(Z)\phi=T\phi\circ Z-Z_E\circ\phi \ ,
\end{equation}
which is a vector field along $\phi$.
The generalized Lie derivative has the form 
$\mathbb{L}(Z)\phi=(\phi,\Lie(Z)\phi)$
and the section $\Lie(Z)\phi\colon M\to V(\Tan \pi)$ is called the
\textbf{Lie derivative of the section $\phi$ by $Z$} and is given as $\displaystyle \Lie(Z)\phi=\Big( x^\mu,\xi^\nu\,\derpar{\phi^i}{x^\nu}-\xi^i\circ\phi\Big)$ when the local expression for $Z_E$ is \eqref{diffsEvector}.
\end{definition}

The definition of this Lie derivative must be in agreement with the Lie derivative of $y^i(x)$ defined by the field transformations $\delta y^i(x)$ in (\ref{liederivative}). 
This is achieved by setting 
$\Lie(Z)\phi^i=\Lie(Z)y^i(x)$
for the Lie derivative of the components $\phi^i$ in \eqref{LieDerivSections1}; 
then, it follows that
the functions $\widetilde{\xi}^i(x)$ in \eqref{liederivative} are given in terms of the component functions $\xi^i(x,y)$ of $Z_E$ and any local section $\phi$ as 
\begin{equation}
\label{xiA}
\widetilde{\xi}^i(x)=\xi^i(x,y)\circ\phi\ .
\end{equation}
For theories in which $y^i(x)$ are scalar fields, diffeomorphisms on the base $M$ produce field variations $\delta y^i(x)$ for which $\widetilde{\xi}^i(x)=0 \ \Longrightarrow \ \xi^i(x,y)=0$.

On the other hand, if the fields are, in general, tensor fields on $M$ of type $(k,l)\not=(0,0)$ (i.e., $T\in\mathfrak{T}^{(k,l)}(\Tan M)$), then the functions $\widetilde{\xi}^i(x)$ are obtained, as usual, from the Jacobian (and its inverse) associated with the coordinate transformation $x^\mu\rightarrow x^\mu+\xi^\mu(x)$. 
That is, if the fields $y^i(x)$ are the components of tensor fields $T\equiv (T^{\mu_1,\dotsc,\mu_r}_{\nu_1,\dotsc,\nu_s}(x))$, then 
\begin{equation}
\label{TensorTransformation}
(T')^{\mu_1,\dotsc,\mu_r}_{\nu_1,\dotsc,\nu_s}(x') = \left(\frac{\partial x'^{\mu_1}}{\partial x^{\alpha_1}}\right) \dotsb \left(\frac{\partial x'^{\mu_r}}{\partial x^{\alpha_r}}\right) \left(\frac{\partial x^{\beta_1}}{\partial x'^{\nu_1}}\right) \dotsb \left(\frac{\partial x^{\beta_s}}{\partial x'^{\nu_s}}\right) T^{\alpha_1,\dotsc,\alpha_r}_{\beta_1,\dotsc,\beta_s}(x)\ .
\end{equation}
The left-hand side of the expression above can be Taylor expanded around $x$ giving the transport term as
$$
(T')^{\mu_1,\dotsc,\mu_r}_{\nu_1,\dotsc,\nu_s} (x') = (T')^{\mu_1,\dotsc,\mu_r}_{\nu_1,\dotsc,\nu_s} (x+\xi)
= (T')^{\mu_1,\dotsc,\mu_r}_{\nu_1,\dotsc,\nu_s} (x) + \xi^\alpha\,\derpar{(T')^{\mu_1,\dotsc,\mu_r}_{\nu_1,\dotsc,\nu_s}(x)}{x^\alpha} \ ,
$$
while performing the same Taylor expansion of the right-hand side of \eqref{TensorTransformation} for each Jacobian (and each inverse Jacobian) gives the global variation of the fields $\Delta T^{\mu_1,\dotsc,\mu_r}_{\nu_1,\dotsc,\nu_s}(x)$ given as
\begin{align}
\label{DeltaT}
\Delta T^{\mu_1,\dotsc,\mu_r}_{\nu_1,\dotsc,\nu_s}= \sum_{\mu=\mu_1}^{\mu_r}\Big(\derpar{\xi^\mu}{x^\lambda}\Big)T^{\mu_1,\dotsc,(\mu\rightarrow\lambda),\dotsc,\mu_r}_{\nu_1,\dotsc,\nu_s} -\sum_{\nu=\nu_1}^{\nu_s}\Big(\derpar{\xi^\lambda}{x^\nu}\Big)T^{\mu_1,\dotsc,\mu_r}_{\nu_1,\dotsc,(\nu\rightarrow \lambda),\dotsc,\nu_s}\ .
\end{align} 
It follows that the local field variation given by the Lie derivative with respect to $\displaystyle Z=-\xi^\mu(x)\frac{\partial}{\partial x^\mu}\in\vf(M)$ is
\beann
\delta T^{\mu_1,\dotsc,\mu_r}_{\nu_1,\dotsc,\nu_s} (x) &=& 
(T')^{\mu_1,\dotsc,\mu_r}_{\nu_1,\dotsc,\nu_s} (x) - T^{\mu_1,\dotsc,\mu_r}_{\nu_1,\dotsc,\nu_s} (x) =  
\Lie(Z)T^{\mu_1,\dotsc,\mu_r}_{\nu_1,\dotsc,\nu_s} (x) \\ &=&
-\xi^\lambda\,\derpar{(T')^{\mu_1,\dotsc,\mu_r}_{\nu_1,\dotsc,\nu_s}(x)}{x^\lambda}+\Delta T^{\mu_1,\dotsc,\mu_r}_{\nu_1,\dotsc,\nu_s}(x)\ .
\eeann
Now recalling the relation \eqref{xiA}, it follows that
\begin{equation}
\label{Tensor_xiE}
Z_E=- \xi^\mu\derpar{}{x^\mu}-\Delta^{\mu_1,\dotsc,\mu_r}_{\nu_1,\dotsc,\nu_s}(x,T)\,\frac{\partial}{\partial T^{\mu_1,\dotsc,\mu_r}_{\nu_1,\dotsc,\nu_s}}\ ,
\end{equation}
where now,
\begin{equation}
\label{DeltaT2}
\Delta T^{\mu_1,\dotsc,\mu_r}_{\nu_1,\dotsc,\nu_s}(x)=\Delta^{\mu_1,\dotsc,\mu_r}_{\nu_1,\dotsc,\nu_s}(x,T)\circ\phi\ .
\end{equation}
It is important to note that, when the configuration manifold $E$ is a tensor bundle over $M$ with coordinates $(x^\mu,T^{\mu_1,\dotsc,\mu_r}_{\nu_1,\dotsc,\nu_s})$,
then \eqref{Tensor_xiE} is precisely the canonical lift of $\displaystyle Z\in\vf(M)$ to $E$ 
which is defined as follows:

\begin{definition}
\label{subida}
\ben
\item
Let $\Phi_M\colon M\to M$ be a diffeomorphism.
The \textbf{canonical lift of $\Phi_M$ to $E$} 
is the diffeomorphism $\Phi_E\colon E\to E$ 
defined as follows: for every $(x,T_x)\in E$
where $T_x\in\mathfrak{T}^{(k,l)}(\Tan_xM)$, define
$\Phi_E(x,T_x):=(\Phi_M(x),{\cal T}\Phi_M(T_x))$,
where ${\cal T}\Phi_M$ denotes the canonical transformation of tensors on $M$ induced by $\Phi_M$.
Thus, $\pi\circ\Phi_E=\Phi_M\circ\pi$.
\item
Let $Z\in\vf (M)$ be the vector field induced by local one-parameter groups of diffeomorphisms of $M$, denoted $\phi_t$.
The \textbf{canonical lift of $Z$ to $E$}
is the vector field $Z_E\in\vf(E)$ induced by
local one-parameter groups of diffeomorphisms  $(\phi_{_E})_t$
which are the canonical lifts 
of $\phi_t$
to the configuration bundle $E$.
\een
\end{definition}

In conclusion, the definition of the Lie derivative \eqref{LieDerivSections1} of the local sections $\phi$ is constructed to always agrees with the Lie derivatives of the fields \eqref{liederivative}, 
where $Z_E$ in \eqref{diffsEvector} is the canonical lift of $Z$ to $E$ and, hence, the relation \eqref{xiA} is always satisfied.

\subsection{Lifting transformations from $E$ to $J^1\pi$}
\label{liftingvf}

Let $\Phi\colon E\to E$ be a diffeomorphism which induces a
diffeomorphism on the base space $M$ as $\Phi_M\colon M\to M$
so $\Phi_M\circ\pi=\pi\circ\Phi$.
Then, the {\sl \textbf{canonical lift}} of $\Phi$ to $J^1\pi$
is the diffeomorphism $j^1\Phi\colon J^1\pi\longrightarrow J^1\pi$
defined as
$$
(j^1\Phi)(\bar y):=j^1(\Phi\circ\phi\circ\Phi_M^{-1})(\Phi_M(x))
\quad ; \quad \mbox{\rm for $\bar y\in J^1\pi$} \ .
$$

Now it is possible to define the canonical lift of $\pi$-projectable
vector fields $Y\in\vf(E)$ to $J^1\pi$ for which there exist $Z\in\vf (M)$ such that
the local flows of $Z$ and $Y$ are $\pi^1$-related.
The {\sl \textbf {canonical lift}} of a $\pi$-projectable vector field 
$Y\in\vf(E)$ to $J^1\pi$ is the vector field $j^1Y\in\vf(J^1\pi)$ 
whose local one-parameter groups of diffeomorphisms are
the canonical lifts of the local
one-parameter groups of diffeomorphisms of $Y$.
If $Y\in\vf(E)$ is the canonical lift of $Z\in\vf(M)$ to $E$
whose expression in local coordinates is given by \eqref{diffsEvector} as
$\dst Y=Z_E=-\xi^\mu(x)\derpar{}{x^\mu}-\xi^i(x,y)\derpar{}{y^i}$,
then the canonical lift of $Y\in\mathfrak{X}(E)$ to $J^1\pi$ is
\begin{equation}
\label{liftJ1}
j^1Z_E=-\xi^\mu\derpar{}{x^\mu}-\xi^i\frac{\partial}{\partial y^i}-\left(\derpar{\xi^i}{x^\mu}-y^i_\nu\derpar{\xi^\nu}{x^\mu}+y^j_\mu\frac{\partial\xi^i}{\partial y^j}\right)\frac{\partial}{\partial y^i_\mu}\in\mathfrak{X}(J^1\pi)\ .
\end{equation}

These canonical lifts are characterized by the property that they leave 
the canonical structures of the jet bundle $J^1\pi$ invariant.
In particular, the {\sl contact module} and, consequently, 
the {\sl canonical endomorphism} under such canonical lifts. 
This canonical lift can be generalized for vector fields on $E$ that are not $\pi$-projectable. 
If $Y$ is not a $\pi$-projectable vector field,
its canonical lift $j^1Y\in\vf(J^1\pi)$
is the only vector field that is $\pi^1$-projectable to $Y$ and leaves  the canonical structures of $J^1\pi$ invariant (see \cite{EMR-96,book:Saunders89} for details).
In local coordinates, a vector field $Y=Z_E\in\vf(E)$ is not $\pi$-projectable if
$\xi^i\equiv\xi^i(x,y)$. Then,
$$
j^1Z_E=-\xi^\mu\derpar{}{x^\mu}-\xi^i\derpar{}{y^i}-\left(\derpar{\xi^i}{x^\mu}-
y^i_\nu\left(\derpar{\xi^\nu}{x^\mu}+
y^j_\mu\derpar{\xi^\nu}{y^j}\right)+y^j_\mu\derpar{\xi^i}{y^j}\right)
\derpar{}{y^i_\mu} \ .
$$
In the same way one can define the canonical lift of any diffeomorphism $\Phi\colon E\to E$ to $J^1\pi$.

The variation of the spacetime derivatives of the fields, 
$\delta y^i_\mu(x)=\delta\left(\derpar{y^i(x)}{x^\mu }\right)=\Lie(Z)\derpar{y^i(x)}{x^\mu}$, can also be characterized by the generalized Lie derivative $\Lie(Z)j^1\phi$ of the first--jet prolongations $j^1\phi:M\rightarrow J^1\pi:x^\mu\mapsto \Big(x^\mu,y^i(x),\derpar{y^i(x)}{x^\mu}\Big)$ with respect to the vector field $\displaystyle Z=-\xi^\mu(x)\frac{\partial}{\partial x^\mu}\in\vf(M)$, where now
\begin{equation}
\label{jetLderiv}
\mathbb{L}(Z)j^1\phi=Tj^1\phi\circ Z-j^1Z\circ j^1\phi\ ,
\end{equation}
and $\mathbb{L}(Z)j^1\phi=(\phi,\Lie(Z)\phi)$.
It thereby follows that
\begin{equation}
\label{LieDerivSections2}
\Lie(Z)(j^1\phi)^i_\mu=\delta y^i_\mu(x)=-\xi^\nu\derpar{y^i_\mu}{x^\nu}-\derpar{\xi^\nu}{x^\mu}\derpar{y^i}{x^\nu}+\derpar{\widetilde{\xi}^i}{x^\mu}+
\derpar{y^j}{x^\mu}\frac{\partial\widetilde{\xi}^i}{\partial y^j}\ ,
\end{equation}
where now $\widetilde{\xi}^i(x)=\xi^i(x,y)\circ j^1\phi$ as before.
Furthermore, $\xi^i=0$ for scalar fields transformed under spacetime diffeomorphisms and the equations above simplify accordingly. 

In the case where the fields are tensor fields $T^{\mu_1,\dotsc,\mu_r}_{\nu_1,\dotsc,\nu_s}(x)$, the canonical lift $j^1Z_E$ in \eqref{liftJ1} of $Z_E$ in \eqref{Tensor_xiE} to $J^1\pi$ is 
\begin{equation*}
j^1Z_E=-\xi^\mu\derpar{}{x^\mu}-\Delta^{\mu_1,\dotsc,\mu_r}_{\nu_1,\dotsc,\nu_s}(x,T)\,\frac{\partial}{\partial T^{\mu_1,\dotsc,\mu_r}_{\nu_1,\dotsc,\nu_s}}-\Gamma^{\mu_1,\dotsc,\mu_r}_{\alpha\nu_1,\dotsc,\nu_s}\,\frac{\partial}{\partial T^{\mu_1,\dotsc,\mu_r}_{\alpha\nu_1,\dotsc,\nu_s}}\ ,
\end{equation*}
where the $\Delta^{\mu_1,\dotsc,\mu_r}_{\nu_1,\dotsc,\nu_s}(x,T)\in C^\infty(E)$ are again given by \eqref{DeltaT} and \eqref{DeltaT2} while

\begin{align*}
\begin{split}
\Gamma^{\mu_1,\dotsc,\mu_r}_{\alpha\nu_1,\dotsc,\nu_s}&= \sum_{\mu=\mu_1}^{\mu_r}\derpar{\xi^\mu}{x^\lambda}\,\derpar{T^{\mu_1,\dotsc,(\mu\rightarrow\lambda),\dotsc,\mu_r}_{\nu_1,\dotsc,\nu_s}}{x^\alpha}  
-\sum_{\nu=\nu_1}^{\nu_s}\derpar{\xi^\lambda}{x^\nu}\,\derpar{T^{\mu_1,\dotsc,\mu_r}_{\nu_1,\dotsc,(\nu\rightarrow \lambda),\dotsc,\nu_s}}{x^\alpha} \\
& +\sum_{\mu=\mu_1}^{\mu_r}\derpar{\xi^\mu}{x^\lambda}\,T^{\nu_1,\dotsc,\nu_s}   -\sum_{\nu=\nu_1}^{\nu_s}\derpar{\xi^\lambda}{x^\nu}\,T^{\mu_1,\dotsc,\mu_r}_{\alpha\nu_1,\dotsc,(\nu\rightarrow \lambda),\dotsc,\nu_s}
-T^{\ \ \mu_1,\dotsc,\mu_r}_{\beta\nu_1,\dotsc,\nu_s}\,\derpar{\xi^\beta}{x^\alpha} \ .
\end{split} 
\end{align*}

\section{Symmetries for Lagrangian and Hamiltonian field theories}
\label{syms2}

The concepts presented in the previous sections will be developed in this section in the 
Lagrangian and De Donder--Weyl Hamiltonian formalisms of first-order field theories.
See \cite{art:deLeon_etal2004,EMR-99b,Gc-73,art:GPR-2016,GR-3,RWZ-2016}
for other details and considerations.

\subsection{First-order Lagrangian field theories}
\label{S2}

In this section, a comprehensive review of the geometric framework for first-order Lagrangian field theories is given. 
For more details, see 
\cite{art:Aldaya_Azcarraga78_2,EMR-96,art:Echeverria_Munoz_Roman98,Gc-73,art:GPR-2016,GS-73,book:Saunders89}.

The phase space for the Lagrangian formulation of a first-order field theories is the so-called {\sl (first-order) multivelocity phase space} which is
the first-order jet bundle $J^1\pi$ of a bundle $\pi\colon E\to M$ called the {\sl configuration bundle} of the theory and the field states $y^i(x)$ are given by local sections $\phi:M\rightarrow E:x^\mu\mapsto (x^\mu,y^i(x))$ as in Section \ref{lifting}.

Each particular field theory is usually specified by a {\sl Lagrangian}.
However, the definition of the Lagrangian differs between the physics and differential geometry communities. 
In the differential geometry literature, the Lagrangian is defined as a $C^\infty$ function on $J^1\pi$ while, in the physics literature, the Lagrangian is defined as a functional on the space of {\sl jet prolongations} $j^1\phi$ from $M$ to $J^1\pi$. 
Both definitions can be used equivalently to formulate the variational principle which gives rise the field equations of the theory under investigation. However, working on $J^1\pi$ has several geometric advantages over working on the space of jet prolongations. 
One advantage is that $J^1\pi$ is a finite dimensional manifold while the space of jet prolongations, which is a space of local sections from $M$ to $J^1\pi$, is an infinite-dimensional manifold.

Thus, a first-order field theory is described by
a first-order Lagrangian density $\Lag \in \df^{m}(J^1\pi)$ which is a $\overline{\pi}^1$-semibasic $m$-form denoted as
$\mathscr{L}=L\,\omega\in\Omega^m(J^1\pi)$ 
and $L\in\Cinfty(J^1\pi)$ is called the {\sl Lagrangian function}.
The Lagrangian phase space $J^1\pi$ possesses a canonical structure called the {\sl canonical endomorphism}
${\cal V}$ which is a $(1,2)$-tensor field whose local expression is given as
 $\displaystyle {\cal V}=\left(\d y^i-y^i_\mu\d x^\mu\right)\otimes
\derpar{}{y^i_\nu}\otimes\derpar{}{x^\nu}$. Using this structure,
the {\sl \textbf{Poincar\'e--Cartan $m$}} and {\sl \textbf{$(m+1)$-forms}}
associated with~$\Lag$ are defined as
$\Theta_{\Lag}:=\inn({\cal V})\d\Lag+\Lag\in\df^{m}(J^1\pi)$ and
$\Omega_{\Lag}:= -\d\Theta_{\Lag}\in\df^{m+1}(J^1\pi)$ respectively 
and they have the following coordinate expressions:
\begin{align*}
\Theta_{\mathscr{L}} &= \frac{\partial L}{\partial y^i_\mu}\d y^i\wedge\d^{m-1}x_\mu -\left(\frac{\partial L}{\partial y^i_\mu}y^i_\mu-L\right)\d^m x \equiv \frac{\partial L}{\partial y^i_\mu}\d y^i\wedge\d^{m-1}x_\mu -E_\Lag\,\d^m x\ ,
\\
\Omega_{\Lag} &=
-\frac{\partial^2L}{\partial y^j_\nu\partial y^i_\mu}
\,\d y^j_\nu\wedge\d y^i\wedge\d^{m-1}x_\mu
-\frac{\partial^2L}{\partial y^j\partial y^i_\mu}\,\d y^j\wedge
\d y^i\wedge\d^{m-1}x_\mu
\nonumber  \\ & \quad
+\frac{\partial^2L}{\partial y^j_\nu\partial y^i_\mu}\,y^i_\mu\,
\d y^j_\nu\wedge\d^mx  +
\left(\frac{\partial^2L}{\partial y^j\partial y^i_\mu}y^i_\mu
 -\derpar{L}{y^j}+\frac{\partial^2L}{\partial x^\mu\partial y^j_\mu}
\right)\d y^j\wedge\d^mx \ .
\end{align*}
The couple $(J^1\pi,\Omega_\Lag)$ is called a {\sl first-order Lagrangian system} 
and it is said to be {\sl regular} if $\Omega_\Lag$ is $1$-nondegenerate
(that is, a {\sl multisymplectic form\/}) and {\sl singular} otherwise
(i.e., $\Omega_\Lag$ is {\sl premultisymplectic\/}).
In the terminology of multisymplectic geometry, $\Theta_\Lag$ is said to be a {\sl (pre)multisymplectic potential} of $\Omega_\Lag$.
The regularity condition is locally equivalent to demanding that
the generalized Hessian matrix
$\displaystyle H_{ij}^{\mu\nu}=\frac{\partial^2L}
{\partial y^i_\mu\partial y^j_\nu}$
be non-singular everywhere on $J^1\pi$.

The solutions to the Lagrangian variational problem stated for a Lagrangian $\Lag$
are holonomic sections $j^1\phi\colon M\to J^1\pi$ such that
\beq
(j^1\phi)^*\inn(X)\Omega_\Lag= 0 \, , \quad \text{for every }X \in \vf(J^1\pi) \ .
\label{eqsec}
\eeq
Equivalently, $j^1\phi$ are the integral sections of a class of locally decomposable
and holonomic multivector fields 
$\{ {\bf X}_{\Lag}\}\subset\vf^m(J^1\pi)$ which satisfy 
\beq
\inn ({\bf X}_{\Lag})\Omega_{\Lag}=0\ .
 \label{lageqmvf}
\eeq
Furthermore, the holonomic multivector fields must be $\bar\pi^1$-transverse, 
\beq
\inn({\bf X}_\Lag)\omega\neq 0 \ .
 \label{fundeqs}
\eeq
This condition can be fixed by taking
a representative in the class $\{{\bf X}_\Lag\}$ such that
$\inn({\bf X}_\Lag)\omega=1$.
The holonomic sections which solve the field equations \eqref{lageqmvf} represent
the physical states of the theory.
In local coordinates, the field equations for sections are given as
$$
 \derpar{L}{y^i}\circ j^1\phi-
\derpar{}{x^\mu}\left(\derpar{L}{y_\mu^i}\circ j^1\phi\right)= 0 \ ,
$$
which are the {\sl Euler--Lagrange equations} of the system.

\begin{remark}
\label{consistency}{\rm
Equation \eqref{lageqmvf} has a multiplicity of solutions
(even in the regular case)
\cite{art:Echeverria_Munoz_Roman98,EMR-99b}.
This means that there is no unique class of integrable multivector fields
(i.e., a unique distribution)  
which solve the field equations \eqref{lageqmvf} on $J^1\pi$. 
Instead, there is a multiplicity of integral sections passing through every point in $J^1\pi$ which solve the field equations.
If the Lagrangian is singular, there is another arbitrariness 
which comes from the degeneracy of the form $\Omega_\Lag$
and is related to the existence of  {\sl geometric gauge symmetries}.
In general, when $(J^1\pi,\Omega_\Lag)$ is a solvable singular Lagrangian system, the field equations admit stable solutions given by multivector fields which are
locally decomposable and $\bar\pi^1$-transverse on some $\bar\pi^1$-transverse submanifold $S_f\subseteq J^1\pi$.
Furthermore, these multivector fields 
are not always integrable (even for regular field theories).
In addition, the multivector field solutions are not necessarily holonomic on all of $S_f$,
but might be instead only on another $\bar\pi^1$-transverse submanifold ${\cal S}_f\subseteq S_f$
to which these holonomic multivector fields solution to field equations must be tangent.
This means that the image of the holonomic sections 
which solve the field equations \eqref{eqsec} are on ${\cal S}_f$ which is called the {\sl final constraint submanifold} on which
the field equations are said to have ``consistent solutions".
The {\sl constraint algorithm} needed to find ${\cal S}_f$
is overviewed in Section \ref{proj}.
See \cite{LMMMR-2005,GGR-2022} for a deeper analysis of these features.
} \end{remark}

\subsection{De Donder--Weyl Hamiltonian Formalism}
\label{hamfor}
This section provides a review of the (pre)multisymplectic De Donder--Weyl formlism for classical field theories.
See \cite{CCI-91,LMM-96,HK-04,art:Roman09} and references therein for more details.

The De Donder--Weyl Hamiltonian formalism for regular Lagrangian (multisymplectic)
field theories is performed on the so-called
{\sl multimomentum bundle} $J^{1*}\pi$
which is constructed as follows:
consider the {\sl extended multimomentum bundle} 
$\Lambda_2^m\Tan^*E$ which is the bundle of $m$-forms on
$E$ vanishing by contraction with two $\pi$-vertical vector fields; then $J^{1*}\pi\equiv\Lambda_2^m\Tan^*E/\Lambda^m_1\Tan^*E$
(where $\Lambda^m_1\Tan^*E$ is the bundle of $\pi$-semibasic $m$-forms on
$E$).
Natural coordinates on $\Lambda_2^m\Tan^*E$ adapted to the bundle $\pi\colon E\to M$ 
are $(x^\nu,y^i,p^\nu_i,p)$ (so $\dim\,\Lambda_2^m\Tan^*E=nm+n+m+1$), and 
natural coordinates on $J^{1*}\pi$ are $(x^\mu,y^i,p_i^\mu)$ (so 
$\dim\, J^{1*}\pi=nm+n+m$).
The natural projections for these bundles are
$$
\bar\tau\colon J^{1*}\pi\to M \,, \quad
\tau\colon J^{1*}\pi\to E \,, \quad
\mathfrak{p}\colon\Lambda_2^m\Tan^*E\to J^{1*}\pi \,.
$$
The {\sl Legendre map} $\mathscr{FL}\colon J^1\pi\to J^{1*}\pi$
associated with a Lagrangian function $L\in\Cinfty(J^1\pi)$ is locally given by 
$$
 \mathscr{FL}^*x^\nu = x^\nu \quad , \quad
 \mathscr{FL}^*y^i = y^i  \quad , \quad
 \mathscr{FL}^*p_i^\nu =\displaystyle\derpar{L}{y^i_\nu} \ .
$$
 The Lagrangian $L$ is regular if, and only if,
 $\mathscr{FL}$ is a local diffeomorphism;
$\mathscr{FL}$ is {\sl hyperregular} when $\mathscr{FL}$ is a global diffeomorphism.
Then, there exist $\Theta_{\rm h}\in\df^m(J^{1*}\pi)$
and $\Omega_{\rm h}:=-\d\widetilde\Theta_{\rm h}\in\df^{m+1}(J^{1*}\pi)$
such that 
$\mathscr{FL}^*\Theta_{\rm h}=\Theta_{\Lag}$
and $\mathscr{FL}^*\Omega_{\rm h}=\Omega_{\Lag}$ 
which are called the {\sl\textbf{Hamilton--Cartan $m$}} and {\sl\textbf{ $(m+1)$-forms}}; $\Omega_{\rm h}$
is the multisymplectic form on $J^{1*}\pi$ and the couple $(J^{1*}\pi,\Omega_{\rm h})$
is the {\sl Hamiltonian system} associated with the (hyper)regular Lagrangian system $(J^1\pi,\Omega_\Lag)$.
The local expressions for $\Theta_{\rm h}$ and $\Omega_{\rm h}$ are
$$
 \Theta_{\rm h}= p_i^\mu\d y^i\wedge\d^{m-1}x_\mu -H\,\d^mx 
\quad  , \quad
\Omega_{\rm h}= -\d p_i^\mu\wedge\d y^i\wedge\d^{m-1}x_\mu +
\d H\wedge\d^mx \ ,
$$
where, 
$$
H=(\mathscr{FL}^{-1})^*E_\Lag=p^\mu_i(\mathscr{FL}^{-1})^*y^i_\mu-(\mathscr{FL}^{-1})^*L\in\Cinfty(J^{1*}\pi)\ ,
$$ 
is the {\sl\textbf{De Donder--Weyl Hamiltonian function}}.
The field equations can be obtained from the so-called {\sl Hamilton--Jacobi variational principle}
and their solutions are sections $\psi\colon M\to J^{1*}\pi$ 
which satisfy
 $$
 \psi^*\inn (X)\Omega_{\rm h}= 0 \quad , \quad \mbox{\rm for every $X\in\vf (J^{1*}\pi)$} \ .
 $$
Equivalently, such sections are integral sections of a class of integrable and
 $\bar\tau$-transverse multivector fields
 $\{ {\bf X}_{\rm h}\}\subset\vf^m(J^{1*}\pi)$ satisfying 
$$
 \inn ({\bf X}_{\rm h})\Omega_{\rm h}=0 \quad , \quad
 \mbox{\rm for every ${\bf X}_{\rm h}\in\{ {\bf X}_{\rm h}$\}}\  .
$$
The $\bar\tau$-transversality condition is fixed (as in the Lagrangian setting) by taking
a representative in the class $\{{\bf X}_{\rm h}\}$ such that
$\inn({\bf X}_{\rm h})\omega=1$.
Working with the natural local coordinates on $J^{1*}\pi$, the local sections are written as $\psi=(x^\mu ,y^i(x^\nu),p^\mu_i(x^\nu))$ and satisfy
$$
 \derpar{(y^i\circ\psi)}{x^\mu}=
 \derpar{H}{p^\mu_i}\circ\psi
\quad ,\quad
 \derpar{(p_i^\mu\circ\psi)}{x^\mu}=
 - \derpar{H}{y^i}\circ\psi \ .
$$
which are the {\sl Hamilton--De Donder--Weyl equations} of the system.

For singular Lagrangians, some minimal conditions must be imposed in order to ensure the existence of a De Donder--Weyl Hamiltonian description.
In particular, one considers the so-called
{\sl almost-regular Lagrangians} $L\in\Cinfty(J^1\pi)$ which are those such that:
\ (i) $P_0:=\mathscr{FL}(J^1\pi)$ is a closed submanifold of $J^{1*}\pi$
which is called the {\sl primary constraint submanifold},
\ (ii) $\mathscr{FL}$ is a submersion onto its image, and
\ (iii) for every $\bar y\in J^1\pi$, the fibers $\mathscr{FL}^{-1}(\mathcal{\mathscr{FL}}(\bar y))$
are connected submanifolds of $J^1\pi$.
Observe that, as $P_0:=\mathscr{FL}(J^1\pi)$, it follows that 
$P_0\to E\to M$.
Then, denoting by $\jmath_0\colon P_0\hookrightarrow J^{1*}\pi$ the natural embedding of the primary constraint submanifold,
the restriction $\mathscr{FL}_0\colon J^1\pi\to P_0$ is the map defined by
$\mathscr{FL}=\jmath_0\circ \mathscr{FL}_0$.
Furthermore, there exist Hamilton--Cartan forms
$\Theta^0_{\rm h}\in\df^m(P_0)$ and $\Omega^0_{\rm h}=-\d\Theta^0_{\rm h}\in\df^{m+1}({\cal P})$
such that $\Theta_\mathscr{L}=\mathscr{FL}_0^{\ *}\,\Theta^0_{\rm h}$ and
$\Omega_\mathscr{L}=\mathscr{FL}_0^{\ *}\,\Omega^0_{\rm h}$
and also a {\sl De Donder--Weyl Hamiltonian function} $H_0\in\Cinfty(P_0)$
such that $E_\mathscr{L}=\mathscr{FL}_0^{\ *}\,H_0$.
Therefore the coordinate expression of the Hamilton--Cartan forms are
$$
\Theta^0_{\rm h}=\jmath_0^{\,*}(p_i^\mu\d y^i\wedge\d^{m-1}x_\mu)-H_0\,\d^mx
\quad , \quad
\Omega^0_{\rm h}=\jmath_0^{\,*}(-\d p_i^\mu\wedge\d y^i\wedge\d^{m-1}x_\mu)+\d H_0\wedge\d^mx \ .
$$
In general, $\Omega_{\rm h}^0$ is a premultisymplectic form
and $(P_0,\Omega^0_{\rm h})$ is the {\sl Hamiltonian system}
associated with the almost-regular Lagrangian system $(J^1\pi,\Omega_\Lag)$.
The Hamilton--de Donder--Weyl equations are stated as in the regular case. 
When the non-regular Hamiltonian system admits stable solutions,
such solutions to the Hamiltonian field equations exist only on a submanifold $P_f\subseteq P_0$
which is obtained by implementing the corresponding constraint algorithm described later in Section \ref{proj}.
The submanifold $P_f$ is the {\sl final constraint submanifold} in the De Donder--Weyl Hamiltonian formalism and
which satisfies $\mathscr{FL}({\cal S}_f)=\mathscr{FL}(S_f)=P_f\subset J^{1*}\pi$
since ${\cal S}_f\hookrightarrow S_f$ is defined on $S_f$ by
non-$\mathscr{FL}$-projectable constraints (see Section \ref{proj}).

\subsection{Symmetries, conserved quantities, and multimomentum maps}

Recall that the Lagrangian formalism takes place on the bundle $J^1\pi\to E\to M$
both in the regular and singular cases.
Then, all the definitions and results from Section \ref{syms} regarding conserved quantities and symmetries apply directly.
The concepts of {\sl Lagrangian conserved quantities} (or {\sl conservation laws}), {\sl Lagrangian (infinitesimal) symmetries, 
Lagrangian (infinitesimal) Noether symmetries, Lagrangian gauge symmetries}, and {\sl Lagrangian multimomentum maps} of a Lagrangian system $(J^1\pi,\Omega_\Lag)$ are also well-defined.

Nevertheless, in this situation there are some new insights to be taken into account,
due to the fact that now, sections, diffeomorphisms and vector fields in $J^1\pi$ 
can be canonical lifts of sections, diffeomorphisms and vector fields in $E$, respectively.
Therefore we define:

\begin{definition}
\label{defsymm1}
\ \ 
\ben
\item
A (Noether) symmetry $\Phi\colon J^1\pi\to J^1\pi$ of a Lagrangian system $(J^1\pi,\Omega_\Lag)$
is said to be \textbf{natural} if $\Phi$ is a canonical lift; i.e., $\Phi=j^1\varphi$ for a diffeormorphism $\varphi\colon E\to E$.
\item
 An infinitesimal (Noether) symmetry $X\in\vf(J^1\pi)$ of a Lagrangian system $(J^1\pi,\Omega_\Lag)$
 is said to be \textbf{natural} if $X$ is a canonical lift; i.e.,
$X=j^1Z_E$ for some $Z_E\in\vf (E)$.
\een
 \end{definition}
 
If $j^1\phi\colon M\to J^1\pi$
is a holonomic solution to the field equations \eqref{eqsec} and
$\Phi=j^1\varphi\in{\rm Diff}(J^1\pi)$ is a natural Noether symmetry, then
\bea
(j^1(\varphi\circ\phi))^*\inn(X)\Omega_\Lag&=&
(j^k\phi)^*(j^k\varphi)^*\inn(X)\Omega_\Lag
\nonumber \\ &=&
(j^1\phi)^*\inn((j^1\varphi)_*^{-1}X)((j^1\varphi)^*\Omega_\Lag)=
(j^1\phi)^*\inn(X')\Omega_\Lag=0 \ ,
\label{holcons2}
\eea
since $X'=\Phi_*^{-1}X\in\vf(J^1\pi)$, $\Phi^*\Omega_\Lag=\Omega_\Lag$, 
and $\Phi^*\Omega_\Lag=0$.
Therefore $j^1(\varphi\circ\phi)$ is also a 
holonomic solution to \eqref{eqsec}
and thus we have proved that:

 \begin{prop}
Every natural (infinitesimal) Noether symmetry transforms holonomic solutions to the Lagrangian field equations into equivalent holonomic solutions.
\label{symsect2}
\end{prop}

As stated for the general case, when $(J^1\pi,\Omega_\Lag)$ is a singular Lagrangian system 
with final constraint submanifold ${\cal S}_f\hookrightarrow J^1\pi$, 
then (Noether) symmetries must be diffeomorphisms leaving ${\cal S}_f$ invariant
and infinitesimal (Noether) symmetries are vector fields tangent to ${\cal S}_f$
and satisfying the conditions of Definition \ref{Cartansym} at least on ${\cal S}_f$.

Furthermore, as stated in Remarks \ref{fiberpreserve} and \ref{fiberpreserve2},
we will be specially interested in the case where (Noether) symmetries
are fiber-preserving diffeomorphisms for the bundle $\bar\pi^1\colon J^1\pi\to M$
and infinitesimal (Noether) symmetries are
$\bar\pi^1$-projectable vector fields.

Similarly, for Lagrangian gauge symmetries we can define:

\begin{definition}
A geometric infinitesimal Lagrangian gauge symmetry (i.e., a Lagrangian gauge vector field) 
$X\in\vf (J^1\pi)$ 
of a Lagrangian system $(J^1\pi,\Omega_\Lag)$ is said to be \textbf{natural} if
$X=j^1Z_E$ for some vector field $Z_E\in\vf(E)$.
\end{definition}

As a consequence of Proposition \ref{symsect2}, if $X\in\vf (J^1\pi)$ is a natural Lagrangian gauge vector field, then it transforms holonomic sections of the projection $\bar\pi^1$
which are solutions to the Lagrangian field equations onto other holonomic sections solutions, all of which represent gauge equivalent physical states.
Nevertheless, it should be noted that a holonomic section could be gauge equivalent to a non-holonomic one (both of them solutions to the field equations)
when they are related by gauge transformations generated by gauge vector fields which are not natural.

In local coordinates, if $X$ is a natural geometric gauge vector field, then the corresponding vector field $Z_E\in\vf(E)$ only has the component
$\displaystyle Z_E=-\xi^i(x,y)\frac{\partial}{\partial y^i}\in\vf(E)$ and it follows that
the gauge vector field $X=j^1Z$ is given by
\begin{equation*}
X=-\xi^i\frac{\partial}{\partial y^i}-\left(\derpar{\xi^i}{x^\mu}+y^j_\mu\frac{\partial\xi^i}{\partial y^j}\right)\frac{\partial}{\partial y^i_\mu}\in \vf(J^1\pi)\ .
\end{equation*}

\parskip=8pt

A particular type of symmetries in the Lagrangian formalism are the following:
\newpage
\begin{definition}
\label{gaugelag}
\ \ 
\ben
\item
A \textbf{Lagrangian symmetry} of a Lagrangian system $(J^1\pi,\Omega_\Lag)$ is a diffeomorphism
$\Phi\colon J^1\pi\to J^1\pi$ that leaves $\Lag$ invariant:
$\Phi^*\Lag=\Lag$.

If $\Phi=j^1\varphi$ for some
fiber-preserving diffeomorphism $\varphi\colon E\to E$, then 
the Lagrangian symmetry is said to be \textbf{natural}.
\item
An \textbf{infinitesimal Lagrangian symmetry} of a Lagrangian system $(J^1\pi,\Omega_\Lag)$ is a
vector field $X\in\vf (J^1\pi)$ that leaves $\Lag$ invariant.

If $X=j^1Z_E$, for some $\pi$-projectable vector field $Z_E\in\vf(E)$, then 
the infinitesimal Lagrangian symmetry is said to be \textbf{natural}.
\label{sym0}
\een
\end{definition}

Observe that, given a diffeomorphism $\varphi\colon E\to E$ and $Z_E\in\vf(E)$, then
\beann
(j^1\varphi)^*\mathscr{L}=\mathscr{L} &\Longleftrightarrow&
(j^1\varphi)^*\Theta_\mathscr{L}=\Theta_\mathscr{L}\ , \\
\Lie(j^1Z_E)\mathscr{L}=0 &\Longleftrightarrow&\Lie(j^1Z_E)\Theta_\mathscr{L}=0\ ,
\eeann
and hence, (infinitesimal) Lagrangian symmetryies are (infinitesimal) exact Noether symmetries.

Nevertheless, a generic (infinitesimal) Lagrangian symmetry
does not necessarily leave the Poincar\'e--Cartan form $\Omega_\Lag$ invariant
unless it is a natural Lagrangian symmetry.
Likewise, as $\Omega_\Lag$ is not
canonical (since it depends on the choice of the Lagrangian density
$\Lag$), it is not invariant under canonical lifts of fiber-preserving diffeomorphisms 
and $\pi$-projectable vector fields unless some additional condition is assumed,
such as the invariance of the Lagrangian density.
In the spirit of this discussion, consider the following type of symmetries:

\begin{definition}
\label{gaugelag2}
\ \ 
\ben
\item
A \textbf{geometric Lagrangian symmetry} of a Lagrangian system $(J^1\pi,\Omega_\Lag)$ is a diffeomorphism
$\Phi\colon J^1\pi\to J^1\pi$ such that:
\ben
\item
$\Phi^*\Lag=\Lag$.
\item
The canonical geometric structures of $J^1\pi$ are invariant by $\Phi$.
\een
\item
An \textbf{infinitesimal geometric Lagrangian symmetry} of a Lagrangian system $(J^1\pi,\Omega_\Lag)$ is a
vector field $X\in\vf (J^1\pi)$ such that:
\ben
\item
 $\Lie(X)\Lag =0$.
 \item
The canonical geometric structures of $J^1\pi$ are invariant
under the action of $X$.
\een
\label{sym}
\een
\end{definition}
It follows that (infinitesimal) natural Lagrangian symmetries are (infinitesimal) geometric Lagrangian symmetries, and
(infinitesimal) geometric Lagrangian symmetries are (infinitesimal) exact Noether symmetries.

It is also worth noting that demanding the invariance of $\Lag$ under some Lie group action on $J^1\pi$ is an exceedingly strong condition that is not necessary in order to obtain the same field equations from the variational principle. 
There are Lie group actions on $J^1\pi$ which produce so-called {\sl\textbf{gauge equivalent Lagrangians}} which give rise to the same Euler-Lagrange equations 
(see, for instance, \cite{first,Krupka}). Gauge equivalent Lagrangian densities differ by an exact differential form: $\Phi^*\Lag=\Lag+\text{d}\beta$. 
Recall that, in the physics literature, the analysis of field theories occurs on the jet prolongations $j^1\phi:M\rightarrow J^1\pi$ so that $\widetilde{\Lag}=\widetilde{L}\omega=(j^1\phi)^*\Lag\in C^\infty(M)$. 
In this setting, gauge equivalent Lagrangians differ by a total derivative $\delta\widetilde{L}=\partial_\mu\widetilde{K}^\mu$ where $\widetilde{K}^\mu=(j^1\phi)^*K^\mu$, for some $K^\mu$. 
Furthermore, recall that, when the field variations are produced by diffeomorphisms of $M$ generated by some vector field $\xi\in\vf(M)$, the variation of the Lagrangian function $\widetilde{L}$ on $j^1\phi$ can be written as a Lie derivative of the local sections $j^1\phi$ with respect to $\xi$ so that 
$$
\delta\widetilde{L}=\derpar{\widetilde{L}}{y^i(x)}\delta y^i(x)+\derpar{\widetilde{L}}{y^i_\mu(x)}\delta y^i_\mu(x)\in C^\infty(M)\ ,
$$ 
using equations \eqref{LieDerivSections1} and \eqref{LieDerivSections2}. Then, the corresponding Lie group action $\Phi$ on $J^1\pi$ is generated by the canonical lift $X_\xi\in\vf(J^1\pi)$ of $\xi\in\vf(M)$ to $J^1\pi$ given by \eqref{liftJ1}. It follows that, for $\Phi^*\Lag=\Lag+\text{d}\beta$,
$$
\text{d}\beta=L(X_\xi)\Lag=-\text{d}(K^\mu+L\xi^\mu)\wedge\text{d}^{m-1}x_\mu\in\Omega^m(J^1\pi)\ .
$$
Similarly, when the field variations are gauge transformations, $\delta y^i(x)=\widetilde{\xi}^i(x)=\phi^*\xi^i(x,y)$, it follows that
$$
\text{d}\beta=L(X_\xi)\Lag=-\text{d}K^\mu\wedge\text{d}^{m-1}x_\mu\in\Omega^m(J^1\pi)\ ,
$$
where now $X_\xi$ is given by \eqref{liftJ1} with $\xi^\mu=0$.

\parskip=8pt

Finally, let $G$ be a {\sl group of Noether symmetries} for the Lagrangian system $(J^1\pi,\Omega_\Lag)$,
and let ${\rm J}_\xi\in\df^{m-1}(U)$, for $U\subset J^1\pi$,
be the corresponding Lagrangian multimomentum map.
Now, consider the situation in which the Noether symmetries are exact and they are associated with diffeomorphisms on the base $M$,
as described in Section \ref{liftME}.
Then, denoting $X_\xi\equiv j^1Z_E$, we have $\Lie(j^1Z_E)\Theta_\mathscr{L}=0$, 
and (modulo exact differential forms) the {\sl\textbf{Lagrangian multimomentum map}} is given as
$$
{\rm J}_\mathscr{L}({X_\xi})=-\inn(X_\xi)\Theta_\mathscr{L}=\derpar{L}{y^i_\mu}\left(\xi^i\text{d}^{m-1}x_\mu-\xi^\nu y^i_\mu\text{d}^{m-1}x_\nu-\xi^\nu\text{d}y^i\wedge\text{d}^{m-2}x_{\mu\nu}\right)+L\,\xi^\nu\text{d}^{m-1}x_\nu\  .
$$
The corresponding Noether current is given by 
$$
{\rm j}^\mu_\xi\text{d}^{m-1}x_\mu=
(j^1\phi)^*\left[-i\left(X_\xi\right)\Theta_\mathscr{L}\right] =\left[\left(\derpar{L}{y^i_\mu}\circ j^1\phi\right)\left(\widetilde{\xi}^i-\xi^\nu\derpar{y^i(x)}{x^\nu}\right)+\xi^\mu\left(L\circ j^1\phi\right)\eta^\mu_{\ \nu}\right]\text{d}^{m-1}x_\mu .
$$
The part of the Noether current shown above which corresponds to the infinitesimal spacetime transformations is linear in $\xi$ and
the {\sl\textbf{canonical energy--momentum tensor}}, denoted as usual as $T^\mu_{\ \nu}$, is defined from the terms contracted with $\xi^\nu$ above:
$$
-\xi^\nu\left[\left(\derpar{L}{y^i_\mu}\circ j^1\phi\right)\derpar{y^i(x)}{x^\nu}-\left(L\circ j^1\phi\right)\eta^\mu_{\ \nu}\right]\text{d}^{m-1}x_\mu
\equiv-\xi^\nu T^\mu_{\ \nu}\text{d}^{m-1}x_\mu\ .
$$

\parskip=8pt

Similarly to the Lagrangian formalism, all the definitions and results introduced in Section \ref{syms}
apply straightforwardly to the De Donder--Weyl Hamiltonian setting by taking $\mathscr{M}\equiv J^{1*}\pi\to E\to M$ for regular field theories
or $\mathscr{M}\equiv P_0\to E\to M$ for almost-regular field theories as discussed in the previous section. 
The concepts of
{\sl Hamiltonian conserved quantities} (or {\sl conservation laws}), {\sl Hamiltonian (infinitesimal) symmetries, Hamiltonian (infinitesimal) Noether symmetries, Hamiltonian gauge symmetries},
and {\sl Hamiltonian multimomentum maps} are well-defined for 
regular and almost-regular Hamiltonian systems
$(J^{1*}\pi,\Omega_{\rm h})$ and $(P_0,\Omega_{\rm h}^0)$.

In particular, the {\sl\textbf{Hamiltonian multimomentum map}} can be obtained by applying the push-forward of the Legendre map $\mathscr{FL}_*$ to ${\rm J}_\mathscr{L}({X_\xi})$ or, equivalently, using the $\mathscr{FL}$-projection of $X_\xi$ and contracting it with the corresponding (pre)multisymplectic form on the image of the Legendre map. When the field theory under investigation is regular, the calculation is straightforward and the Hamiltonian multimomentum map is given as
$$
\mathscr{FL}_* \Big({\rm J}_\mathscr{L}({X_\xi})\Big)={\rm J}_h({\mathscr{FL}_* X_\xi})=-i({\mathscr{FL}_* X_\xi})\Omega_h
= p_i^\mu(\xi^i\text{d}^{m-1}x_\mu-\xi^\nu\text{d}y^i\wedge\text{d}^{m-2}x_{\mu\nu})-H\xi^\mu\text{d}^{m-1}x_\mu,
$$
where $H$ is the De Donder--Weyl Hamiltonian function. However, when the field theory is singular, important subtleties (which are detailed below) arise when projecting vector fields via $\mathscr{FL}$.


\subsection{Symmetries in the presence of constraints}
\label{proj}

The projection of Lagrangian symmetries to the Hamiltonian framework is straightforward when the Lagrangian is regular; that is, when the Legendre map $\mathscr{FL}$ is a (local) diffeomorphism.
Given a vector field $X\in\vf(J^1\pi)$ which generates a Lagrangian Noether symmetry on $J^1\pi$; that is, $\Lie(X)\Omega_\mathscr{L}=0$,
the corresponding Hamiltonian symmetry on $J^{1*}\pi$ is generated by the vector field obtained from the push-forward by the Legendre map, $Y=\mathscr{FL}_*X\in\vf(J^{1*}\pi)$;
in fact,
$$
0=\Lie(X)\Omega_\mathscr{L}=\Lie(X)(\mathscr{FL}^*\Omega_{\rm h})=\mathscr {FL}^*[\Lie(Y)\Omega_{\rm h}] \quad \Longleftrightarrow \quad \Lie(Y)\Omega_{\rm h}=0\ ,
$$
where the push-forward by the Legendre map $\mathscr{FL}_*$ is given by the matrix
$$
\Tan{{\mathscr F}\Lag}\equiv
\left(\begin{matrix}
({\rm Id})_{m\times m} & (0)_{m\times n} &  (0)_{m\times nm}
\\
(0)_{n\times m} & ({\rm Id})_{n\times n}& (0)_{n\times nm} 
\\
(0)_{nm\times m} & \left(\displaystyle\frac{\partial^2\Lag}{\partial y^B\partial y^A_\mu} \right) & \left(\displaystyle\frac{\partial^2\Lag}{\partial y^B_\nu \partial y^A_\mu}\right)
\end{matrix}\right) \ .
$$

When the Lagrangian is singular (in particular, almost-regular) the primary constraint submanifold
$P_0=\text{Im}\mathscr{FL}\subset J^{1*}\pi$, is defined by some independent functions $\varphi_I\in C^\infty(J^{1*}\pi)$. It follows that the null vectors of the Hessian matrix are given by \cite{GGR-2022}:
\begin{equation*}
\left(\gamma^i_\mu\right)_I=\mathscr{FL}^*\derpar{\varphi_I}{p^i_\mu}\ ,
\end{equation*}
and since $\text{ker}\mathscr{FL}_*\subset\vf^{V(\pi)}(J^1\pi)$, the vector fields 
$\Gamma_I\in\text{ker}\mathscr{FL}_*$ can be written using a local basis for $\text{ker}\mathscr{FL}_*$ in natural coordinates on $J^1\pi$ as
$$
\Gamma_I=\left(\gamma^i_\mu\right)_I\derpar{}{y^i_\mu} \ .
$$
Furthermore, since
$$
\text{ker}\, \mathscr{FL}_*=\text{ker}\, \Omega_\mathscr{L}\cap\vf^{{\rm V}(\pi_1)}(J^1E)
\subset\text{ker}\,\Omega_\mathscr{L}\cap\vf^{{\rm V}(\bar{\pi}_1)}(J^1E)=\text{ker}\,\Omega_\mathscr{L}\cap\text{ker}\, \omega ,
$$
it follows that $\Gamma_I\in\text{ker}\,\Omega_\mathscr{L}\cap\text{ker}\, \omega$ and thereby generate geometric gauge symmetries.

In general, the Lagrangian and the Hamiltonian field equations have consistent solutions on the final constraint submanifolds 
${\cal S}_f\subseteq J^1\pi$ and $P_f\subseteq P_o\subset J^{1*}\pi$
respectively, where $\mathscr{FL}({\cal S}_f)=P_f$. 
Moreover, vector fields $X\in\vf(J^1\pi)$ which generate Noether symmetries on $J^1\pi$ must be tangent to ${\cal S}_f$ in order to preserve the full constraint structure of the field theory under investigation;
the tangency of $X$ to ${\cal S}_f$ is guaranteed by ensuring $\Lie(X)\Omega_\mathscr{L}=0$ (at least on ${\cal S}_f$). 

If $X$ is $\mathscr{FL}$-projectable from $J^1\pi$, then,
proceeding similarly as in the regular case, the vector fields which generate
Hamiltonian Noether symmetries on $P_0$ are given by $Y_0=(\mathscr{FL}_0)_*X\in\vf(P_0)$ and it follows that,
$$
0=\Lie(X)\Omega_\mathscr{L}=\Lie(X)(\mathscr{FL}_0^*\Omega_{\rm h}^0)=\mathscr{FL}_0^*[\Lie(Y_0)\Omega_{\rm h}^0] \quad \Longleftrightarrow \quad \Lie(Y_0)\Omega_{\rm h}=0\ ,
$$
since $\mathscr{FL}$ is a submersion. Furthermore, as $\mathscr{FL}({\cal S}_f)=P_f$,
the tangency of $X$ to the Lagrangian final constraint submanifold ${\cal S}_f$
guarantees the tangency of $Y_0$ to the Hamiltonian final constraint submanifold $P_f$, as desired.

However, it is not always the case that a vector field on $J^1\pi$ is projectable onto $P_0$ via $\mathscr{FL}_0$.
As it is well-known, the necessary and sufficient condition is that
$[X,Y]\subset \ker\,\mathscr{FL}_*$, for every $Y\in\ker\,\mathscr{FL}_*$.
Recall that a function $f\in C^\infty(J^1\pi)$ is $\mathscr{FL}$-projectable if, and only if, $\Lie(\Gamma_I)f=0$.
It follows that, locally, a vector field $X\in J^1\pi$ 
can be split as $X=X^o+Y$, where $Y\in\ker\,\mathscr{FL}_*$;
then $X$ is $\mathscr{FL}$-projectable if, and only if, 
the component functions of $X^o$
are $\mathscr{FL}$-projectable functions on $J^1\pi$. 
Furthermore, it is sometimes the case that a vector field on $J^1\pi$ is not $\mathscr{FL}$-projectable as a result of a dependence of some components of $X^o$ on the functions which define Lagrangian constraints on $J^1\pi$.
When this occurs, it is natural to project such a vector field to $J^{1*}\pi$ from the constraint submanifold defined by the aforementioned nonprojectable constraint functions.
It is also important to note that {\sc sopde} Lagrangian constraints are not $\mathscr{FL}$-projectable
(see, for instance, \cite{AGR-2022}). 

The following (commutative) diagram depicts the projection of Lagrangian constraint submanifolds of $J^1\pi$ to Hamiltonian constraint submanifolds of $J^{1*}\pi$:
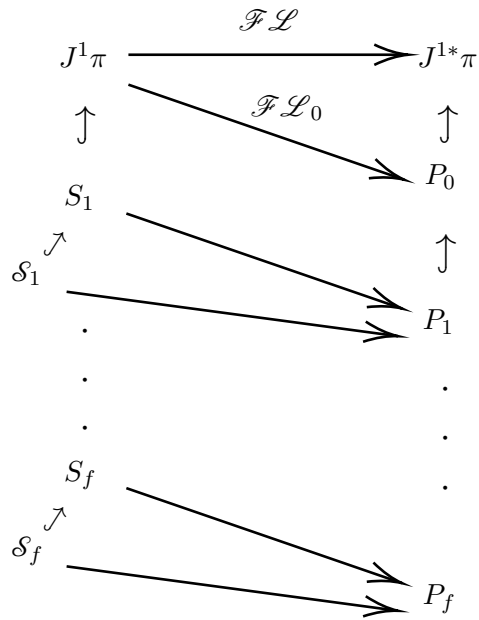
\begin{center}
\tikzset{every picture/.style={line width=0.75pt}} 
\begin{tikzpicture}[x=0.75pt,y=0.75pt,yscale=-1,xscale=1]
\draw [line width=1]    (277.75,3049.51) -- (415.8,3049.51) ;
\draw [shift={(418.8,3049.51)}, rotate = 180] [color={rgb, 255:red, 0; green, 0; blue, 0 }  ][line width=1]    (17.05,-5.13) .. controls (10.84,-2.18) and (5.16,-0.47) .. (0,0) .. controls (5.16,0.47) and (10.84,2.18) .. (17.05,5.13)   ;
\draw [line width=1]    (277.75,3064.61) -- (413.96,3111.6) ;
\draw [shift={(416.8,3112.58)}, rotate = 199.03] [color={rgb, 255:red, 0; green, 0; blue, 0 }  ][line width=1]    (17.05,-5.13) .. controls (10.84,-2.18) and (5.16,-0.47) .. (0,0) .. controls (5.16,0.47) and (10.84,2.18) .. (17.05,5.13)   ;
\draw [line width=1]    (246.57,3169.14) -- (410.71,3191.16) ;
\draw [shift={(413.68,3191.56)}, rotate = 187.64] [color={rgb, 255:red, 0; green, 0; blue, 0 }  ][line width=1]    (17.05,-5.13) .. controls (10.84,-2.18) and (5.16,-0.47) .. (0,0) .. controls (5.16,0.47) and (10.84,2.18) .. (17.05,5.13)   ;
\draw [line width=1]    (276.71,3129.65) -- (412.92,3176.65) ;
\draw [shift={(415.76,3177.62)}, rotate = 199.03] [color={rgb, 255:red, 0; green, 0; blue, 0 }  ][line width=1]    (17.05,-5.13) .. controls (10.84,-2.18) and (5.16,-0.47) .. (0,0) .. controls (5.16,0.47) and (10.84,2.18) .. (17.05,5.13)   ;
\draw [line width=1]    (246.57,3307.75) -- (410.71,3329.77) ;
\draw [shift={(413.68,3330.17)}, rotate = 187.64] [color={rgb, 255:red, 0; green, 0; blue, 0 }  ][line width=1]    (17.05,-5.13) .. controls (10.84,-2.18) and (5.16,-0.47) .. (0,0) .. controls (5.16,0.47) and (10.84,2.18) .. (17.05,5.13)   ;
\draw [line width=1]    (276.71,3268.26) -- (412.92,3315.25) ;
\draw [shift={(415.76,3316.23)}, rotate = 199.03] [color={rgb, 255:red, 0; green, 0; blue, 0 }  ][line width=1]    (17.05,-5.13) .. controls (10.84,-2.18) and (5.16,-0.47) .. (0,0) .. controls (5.16,0.47) and (10.84,2.18) .. (17.05,5.13);
\draw (231.96,3289.76) node [anchor=north west][inner sep=0.75pt]  [rotate=-302.27]  {$\hookrightarrow $};
\draw (243.91,3252.94) node [anchor=north west][inner sep=0.75pt]    {$S_{f}\mathbf{\ }$};
\draw (216.93,3291.27) node [anchor=north west][inner sep=0.75pt]    {$\mathcal{S}_{f}\mathbf{\ }$};
\draw (331.79,3025.5) node [anchor=north west][inner sep=0.75pt]    {$\mathscr{FL}$};
\draw (422.06,3041.81) node [anchor=north west][inner sep=0.75pt]    {$J^{1*} \pi\ $};
\draw (241.13,3041.79) node [anchor=north west][inner sep=0.75pt]    {$J^{1} \pi \ $};
\draw (337.77,3068.67) node [anchor=north west][inner sep=0.75pt]    {$\mathscr{FL}_{0}$};
\draw (425.8,3103.48) node [anchor=north west][inner sep=0.75pt]    {$P_{0}\mathbf{\ }$};
\draw (433.43,3096.79) node [anchor=north west][inner sep=0.75pt]  [font=\Large,rotate=-270.14]  {$\hookrightarrow $};
\draw (231.96,3150.76) node [anchor=north west][inner sep=0.75pt]  [rotate=-302.27]  {$\hookrightarrow $};
\draw (250.52,3097.24) node [anchor=north west][inner sep=0.75pt]  [font=\Large,rotate=-270.14]  {$\hookrightarrow $};
\draw (243.91,3113.94) node [anchor=north west][inner sep=0.75pt]    {$S_{1}\mathbf{\ }$};
\draw (216.93,3152.27) node [anchor=north west][inner sep=0.75pt]    {$\mathcal{S}_{1}\mathbf{\ }$};
\draw (250.52,3193.08) node [anchor=north west][inner sep=0.75pt]  [font=\Large,rotate=-270.14]  {$\cdot $};
\draw (424.76,3176.66) node [anchor=north west][inner sep=0.75pt]    {$P_{1}\mathbf{\ }$};
\draw (432.43,3161.96) node [anchor=north west][inner sep=0.75pt]  [font=\Large,rotate=-270.14]  {$\hookrightarrow $};
\draw (250.52,3217.47) node [anchor=north west][inner sep=0.75pt]  [font=\Large,rotate=-270.14]  {$\cdot $};
\draw (250.52,3241.87) node [anchor=north west][inner sep=0.75pt]  [font=\Large,rotate=-270.14]  {$\cdot $};
\draw (432.39,3222.38) node [anchor=north west][inner sep=0.75pt]  [font=\Large,rotate=-270.14]  {$\cdot $};
\draw (432.39,3246.77) node [anchor=north west][inner sep=0.75pt]  [font=\Large,rotate=-270.14]  {$\cdot $};
\draw (432.39,3272.17) node [anchor=north west][inner sep=0.75pt]  [font=\Large,rotate=-270.14]  {$\cdot $};
\draw (424.78,3315.26) node [anchor=north west][inner sep=0.75pt]    {$P_{f}\mathbf{\ }$};
\end{tikzpicture}
\end{center}

On the left-hand side of the diagram, the Lagrangian constraint algorithm starts with compatibility constraints (if they exist) which define the constraint submanifold $S_1$. 
The next step in the algorithm is to impose the {\sc sopde} condition which may produce additional constraints which define the constraint submanifold $\mathcal{S}_1$. 
The rest of the algorithm continues by imposing tangency of the multivector fields which are solutions to the field equations to each constraint submanifold that appears, 
which may produce additional constraint submanifolds until imposing tangency produces no new constraints (at which point the constraint algorithm  terminates). 
If  no {\sc sopde} constraints arise at the second stage of the algorithm, then the tangecy constraint submanifolds are $S_2$, ..., $S_f$; 
if {\sc sopde} constraints do arise at the second stage of the algorithm, then the tangecy constraint submanifolds are $\mathcal{S}_2$, ..., $\mathcal{S}_f$.

On the right-hand side of the diagram, $P_0$ is the primary constraint submanifold produced by the Legendre map $\mathscr{FL}$ 
while $P_1$ is defined by the compatibility constraints (if they exist) and $P_2$, ..., $P_f$ are the constraint submanifolds produced by the tangency condition. 
Since {\sc sopde} contraints are not $\mathscr{FL}$-projectable, the Lagrangian constraint submanifolds $S_i$ and $\mathcal{S}_i$ 
land on the same Hamiltonian constraint submanifold $P_i$ when acted upon by the Legendre map. 

As mentioned earlier, a vector field  which produces a Noether symmetry on $J^1\pi$ may be $\mathscr{FL}$-projectable only from some of the Lagrangian constraint submanifolds $S_i$ (or $\mathcal{S}_i$). 
If this is the case, then the vector field produced by the push-forward of the Legendre map is a vector field on the corresponding Hamiltonian constraint submanifold $P_i$. 
Furthermore, a Noether symmetry may exist only on one of the Lagrangian constraint submanifolds $S_i$ (or $\mathcal{S}_i$). 
If this is the case, then it follows that, on the Hamiltonian side, the Noether symmetry exists only on the corresponding constraint submanifold $P_i$. 
The collection of different scenarios of how symmetries behave in the presence of premultisymplectic constraints are listed as follows:
\ben
\item
There exists some $X\in\vf(J^1\pi)$ such that $L(X)\Omega_\mathscr{L}=0$ and is $\mathscr{FL}$-projectable only from the constraint submanifold $\mathcal{S}_f\hookrightarrow J^1\pi$. 
Then, on the corresponding Hamiltonian constraint submanifold $P_f\subset P_0$, there exists the vector field $Y=\mathscr{FL}_*X\vert_{\mathcal{S}_f}\in\vf(P_f)$ 
such that $L(Y)\Omega^0_h\vert_{P_f}=0$.
Furthermore, the vector field $X\in\vf(J^1\pi)$ may or may not be the local extension of some $Z\vert_{\mathcal{S}_f}$ to $J^1\pi$ for some $Z\in\vf(J^1\pi)$.
\item
There exists some $X\in\underline{\vf(\mathcal{S}_f)}$ such that $L(X)\Omega_\mathscr{L}\vert_{\mathcal{S}_f}=0$ and $X$ is $\mathscr{FL}$-projectable only from $\mathcal{S}_f$. 
Then, on the corresponding Hamiltonian constraint submanifold $P_f\subset P_0$, 
there exists the vector field $Y\vert_{P_f}=\mathscr{FL}_*X\vert_{\mathcal{S}_f}\in\underline{\vf(P_f)}\vert_{P_f}$ 
such that $L(Y)\Omega^0_{\rm h}\vert_{P_f}=0$.
Furthermore, it is possible to construct a local extension of $Y$ to $P_0$, denoted as $\widetilde{Y}\in\vf(P_0)$; but  $X\in\vf(J^1\pi)$ may or may not be the local extension of some $Z\vert_{\mathcal{S}_f}$ to $J^1\pi$ for some $Z\in\vf(J^1\pi)$.
\een
In the following sections, various field theories are worked out as examples whose symmetry structures, in the presence of constraints, exhibit special cases of the scenarios described above.

\section{Some examples}
\label{examples}

In the next Sections we summarize the main results about gauge and Noether symmetries and their associated conserved quantities (given by the corresponding multimomentum maps)
of some differents and known classical field theories in theoretical physics.
For the details and calculations of the systems studied in Sections
\ref{Ex2}, \ref{Ex4}, and \ref{Ex5} see \cite{GGR-2022}.

\subsection{Bosonic string theories}
\label{Ex2}

Now spacetime $M$ is a smooth $(d+1)$-dimensional manifold endowed with a spacetime metric $G_{\mu\nu}$ with signature $(-+\dotsb +)$.
The string worldsheet $\Sigma$ is a smooth $2$-dimensional manifold and has local coordinates $\sigma^a$ with  $a=0,1$. 
The fields $x^\mu(\sigma)$ are the scalar fields on $\Sigma$ given by the embedding maps $X: \Sigma \rightarrow M: \sigma^a\mapsto x^{\mu}(\sigma)$ (see \cite{BBS} for a thorough presentation and discussion on String Theory);
so the configuration bundle $E$ over $\Sigma$ is $E=\Sigma\times M$, with natural projection $\pi:E\rightarrow \Sigma$ whose sections of are $\phi:\Sigma\rightarrow\Sigma\times M: \sigma^a \mapsto (\sigma^a,x^{\mu}(\sigma))$. 
The Lagrangian and Hamiltonian phase bundles $J^1\pi$ and $J^1\pi^*$ have local coordinates $(\sigma^a,x^\mu,x_a^\mu)$ and $(\sigma^a,x^\mu,p^a_\mu)$ respectively,
and the jet prolongations are $j^1\phi:\Sigma\times M\rightarrow J^1\pi:\sigma^a\mapsto \Big(\sigma^a,x^\mu(\sigma), \derpar{x^\mu}{\sigma^a}(\sigma)\Big)$.

The bosonic string theory is described by the standard {\sl Nambu-Goto Lagrangian density} 
\begin{equation*}
\mathscr{L}=L(\sigma^a, x^{\mu}, x^{\mu}_a) \text{d}^2\sigma= -T \sqrt{-\text{det} g} \ \text{d}^2\sigma= -T\sqrt{-\text{det}(G_{\mu\nu}x^{\mu}_a x^{\nu}_b)} \ \text{d}^2\sigma \ ,
\end{equation*}
where $T$ is called the {\sl string tension}.
This Lagrangian is regular as it can be seen directly from the regularity of the multi-Hessian matrix
\begin{equation}
\label{pbraneHessian}
\frac{\partial^2L}{\partial x_a^\mu \partial x_b^\nu}=-T\sqrt{-\text{det}g}\left[ G_{\mu\nu}g^{ba} -G_{\mu\alpha}G_{\rho\nu}x^\alpha_c x^\rho_i\left(g^{ba}g^{ci}+g^{cb}g^{ai}-g^{ca}g^{bi}\right)\right]\ ,
\end{equation}
where $\displaystyle g^{ba}\equiv(g^{-1})^{ba}= \frac{1}{\text{det}g}\epsilon^{bc}\epsilon^{ad}g_{dc}$.
The Lagrangian energy function and the Poincar\'e--Cartan forms on $J^1\pi$ are
\beann
E_{\mathscr{L}}&=& \frac{\partial L}{\partial x_a^\mu}x_a^\mu - L=-T \sqrt{-\text{det} g}\ (g^{ba}g_{ab}-1)=-T \sqrt{-\text{det} g}\ , \\
\Theta_\mathscr{L}&=&
\frac{\partial L}{\partial x_a^\mu} \ \text{d}x^\mu\wedge \text{d}^{1}\sigma_a-E_\mathscr{L} \wedge\text{d}^2\sigma=
-T\sqrt{-\text{det}g} \left[G_{\mu\nu}g^{ba}x_b^\nu\text{d}x^\mu\wedge\text{d}^1\sigma_a-\text{d}^2\sigma\right] \ , 
\\
\Omega_\mathscr{L}&=&
T\Bigg\{\sqrt{-\text{det}g}\left[ G_{\mu\nu}g^{ba} -G_{\mu\alpha}G_{\beta\nu}x^\alpha_c x^\beta_i\left(g^{ba}g^{ci}+g^{cb}g^{ai}-g^{ca}g^{bi}\right)\right] \text{d}x^b_\nu\wedge\text{d}x^\mu\wedge\text{d}^1\sigma_a\\
& &+\derpar{}{x^\rho}\left(\sqrt{-\text{det}g}\, G_{\mu\nu}g^{ba}x^\nu_b\right)\text{d}x^\rho\wedge\text{d}x^\mu\wedge\text{d}^1\sigma_a\\
& &- \sqrt{-\text{det}g}\, \left[G_{\mu\nu}g^{ba} -G_{\mu\alpha}G_{\beta\nu}x^\alpha_c x^\beta_i\left(g^{ba}g^{ci}+g^{cb}g^{ai}-g^{ca}g^{bi}\right) \right]x^\mu_a\text{d}x^\nu_b\wedge\text{d}^2\sigma \\ 
& & -\left[
\derpar{}{\sigma^a}\left(\sqrt{-\text{det}g}\, G_{\mu\nu}g^{ba}x^\nu_b\right)
+\frac{1}{2}\sqrt{-\text{det}g}\,g^{ba} x^\alpha_ax^\beta_b\partial_{\mu} G_{\alpha\beta}
\right]\text{d}x^\mu\wedge\text{d}^2\sigma\Bigg\} \ .
\eeann
Writing $h_{ab}=(j^1\phi)^*g_{ab}$, the resulting Euler--Lagrange equations are
\beann
& &-\sqrt{-\text{det}h}\, h^{ba}\derpar{x^\alpha}{\sigma^a}\derpar{x^\beta}{\sigma^b}\derpar{G_{\alpha\beta}}{x^\mu}+\derpar{}{\sigma^a}\left(\sqrt{-\text{det}h}\, G_{\mu\nu}h^{ba}\derpar{x^\nu}{\sigma^b}\right)+\derpar{}{x^\rho}\left(\sqrt{-\text{det}h}\, G_{\mu\nu}h^{ba}\derpar{x^\nu}{\sigma^b}\right)\derpar{x^\rho}{\sigma^a}
\\ 
& &+\sqrt{-\text{det}h}\left[ G_{\mu\nu}h^{ba} -G_{\mu\alpha}G_{\beta\nu}\derpar{x^\alpha}{\sigma^c}\derpar{x^\beta}{\sigma^i}\left(h^{ba}h^{ci}+h^{cb}h^{ai}-h^{ca}h^{bi}\right)\right]\frac{\partial^2x^\nu}{\partial\sigma^a\partial\sigma^b}=0 \ .
\eeann

The Legendre map $\mathscr{FL}:J^1\pi\rightarrow J^1\pi^*$ gives
$$
\mathscr{FL}^*\sigma^a=\sigma^a \quad , \quad
\mathscr{FL}^*x^\mu=x^\mu \quad , \quad
\mathscr{FL}^*p^a_{\mu}=-T \sqrt{-\text{det} g}\ G_{\mu\nu}g^{ba}x_b^{\nu} \ ,
$$
which is invertible due to the regularity of the generalized Hessian (\ref{pbraneHessian}). 
Using that
\begin{equation*}
\Pi^{ab}\equiv G^{\mu\nu}p^a_\mu p^b_\nu\ 
\Rightarrow \ \mathscr{FL}^*\Pi^{ab}=-T^2 \text{det}g\ g^{ba} \
\Longleftrightarrow \ \mathscr{FL}^*\text{det}\Pi=(-T^2\text{det}g)^2\text{det}(g^{-1})=T^4\text{det}g \ ,
\end{equation*}
and that
$\displaystyle \Pi_{ab}=\frac{1}{\text{det}\Pi}\epsilon_{cb}\epsilon_{da}\Pi^{cd}$,
it follows that
\begin{equation*}
(\mathscr{FL}^{-1})^*x_b^\nu=-\frac{1}{T}\sqrt{-\text{det}\Pi}\ G^{\mu\nu}\Pi_{ab} p^a_\mu \ ,
\end{equation*}
and therefore the De Donder--Weyl Hamiltonian function is
$$
H(\sigma^a,x^{\mu},p^a_{\mu})\equiv p^a_{\mu}(\mathscr{FL}^{-1})^* x_a^{\mu} - (\mathscr{FL}^{-1})^*L=
 -\frac{1}{T}\sqrt{-\text{det}\Pi}
\in C^\infty(J^1\pi^*) \ .
$$
Then, the Hamilton--Cartan forms on $J^{1*}\pi$ are given as
\begin{subequations}
\beann
\Theta_{\rm h}&=&p^a_{\mu} \text{d}x^{\mu}\wedge \text{d}^1\sigma_a - H\wedge\text{d}^2\sigma=p^a_\mu\text{d}x^\mu\wedge\text{d}^1\sigma_a+\frac{1}{T}\sqrt{-\text{det}\Pi}\ \text{d}^2\sigma \label{eq:stringOmegHa}\ , \\
\Omega_{\rm h}&=&
-\text{d}p^a_{\mu}\wedge \text{d}x^{\mu}\wedge \text{d}^1\sigma_a+\text{d}H\wedge\text{d}^2\sigma \\
&=&-\text{d}p^a_{\mu}\wedge \text{d}x^{\mu}\wedge \text{d}^1\sigma_a-\frac{\sqrt{-\text{det}\Pi}}{T} \Pi_{ba}\left(\frac{1}{2}\partial_\mu G^{\rho\sigma}p^a_\rho p^b_\sigma \text{d}x^\mu + G^{\mu\nu}p^b_\nu\text{d}p^a_\mu\right)\wedge\text{d}^2\sigma \ .
\eeann
\end{subequations}
The field equations are the Hamilton--De Donder--Weyl equations given by
$$
\derpar{p^a_{\mu}}{\sigma^a}-\frac{\sqrt{-\text{det}\Pi}}{2T}\Pi_{ba} \partial_\mu G^{\rho\sigma}p^a_\rho p^b_\sigma=0
\quad , \quad
\derpar{x^\mu}{\sigma^a}+\frac{\sqrt{-\text{det}\Pi}}{T}\Pi_{ba}G^{\mu\nu}p^b_\nu=0\ . 
$$

The worldsheet diffeomorphisms are produced by transformations $\sigma^a+\xi^a(\sigma)$;
so the vector field $\xi\in\vf(\Sigma)$ which generates the worldsheet diffeomorphisms is given by
\begin{equation*}
\xi=-\xi^a\derpar{}{\sigma^a}\in\vf(\Sigma)\ .
\end{equation*}
Then, $\xi\in\mathfrak{X}(\Sigma)$ can be lifted from $\Sigma$ to the trivial bundle $E=\Sigma\times M$ and from $E$ to $J^1\pi$ as
\begin{equation*}
\xi_E=-\xi^a\frac{\partial}{\partial\sigma^a}\in\mathfrak{X}(E)
\quad , \quad
X_\xi=-\xi^a\frac{\partial}{\partial\sigma^a}+x_b^\mu\derpar{\xi^b}{\sigma^a}\frac{\partial}{\partial x_a^\mu} \ \in\mathfrak{X}(J^1\pi) \ .
\end{equation*}
It follows that the field variation given by the generalized Lie derivative (\ref{LieDerivSections1}) of the local sections $\phi:\Sigma\rightarrow E$ with respect to $\xi$ is
$\displaystyle\delta X^\mu(\sigma)=-\xi^a\derpar{X^\mu}{\sigma^a}$.
Furthermore, 
$$
\Lie(X_\xi)\mathscr{L}= 0 \ \Rightarrow\ \Lie(X_\xi)\Theta_\mathscr{L}=0 \ ,
$$
so the multimomentum map on $J^1\pi$ is
\begin{equation*}
{\rm J}_\mathscr{L}(X_\xi)=-i(X_\xi)\Theta_\mathscr{L}=-T\sqrt{-\text{det}g}\left( \epsilon_{ac}\xi^cG_{\mu\nu}g^{ba}x_b^\nu\text{d}x^\mu+\xi^a\text{d}^1\sigma_a\right) \ \in \df^1(J^1\pi) \ .
\end{equation*}
The vector field $Y_\xi=\mathscr{FL}_*X_\xi\in\vf(J^1\pi^*)$ which generates the worldsheet diffeomorphisms on $J^{1*}\pi$ takes the form
\begin{equation*}
Y_\xi=-\xi^a\frac{\partial}{\partial \sigma^a}-\frac{1}{T}\sqrt{-\text{det}\Pi}\ G^{\mu\nu}\Pi_{bc}p^b_\nu\derpar{\xi^c}{\sigma_a}\frac{\partial}{\partial p^\mu_a}\ ,
\end{equation*}
and the corresponding multimomentum map is 
\begin{equation*}
{\rm J}_{\rm h}(Y_\xi)=-i(Y_\xi)\Theta_{\rm h}=\epsilon_{ac}\xi^c p^a_\mu\text{d}x^\mu-\frac{1}{T}\sqrt{-\text{det}\Pi}\ \xi^a\text{d}^1\sigma_a\in\df^1(J^{1*}\pi)\ .
\end{equation*}

On the other hand, spacetime diffeomorphisms are generated by the vector field
\begin{equation*}
\zeta=-\zeta^\mu\frac{\partial}{\partial x^\mu}\in(M) \ .
\end{equation*}
whose canonical lift from $M$ to $E$ and from $E$ to $J^1\pi$ are
\begin{equation*}
\zeta_E=-\zeta^\mu\frac{\partial}{\partial x^\mu}\in\vf(E) \quad , \quad
X_\zeta=-\zeta^\mu\frac{\partial}{\partial x^\mu}-x^\nu_a\partial_\nu\zeta^\mu\frac{\partial}{\partial x^\mu_a}\in\mathfrak{X}(J^1\pi)\ .
\end{equation*}
Then, when $\zeta$ is a Killing vector field on $M$, it follows that
$$
\Lie(X_\zeta)\mathscr{L}=\frac{T}{2}\sqrt{-\text{det}g}\ g^{ba}x^\nu_bx^\mu_a\Lie(\zeta) G_{\mu\nu}\text{d}^2\sigma\zeta_E=-\zeta^\mu\frac{\partial}{\partial x^\mu}\in{E}=0 \ \Rightarrow\  \Lie(X_\zeta)\Theta_\mathscr{L}=0 \ ,
$$
and this means that spacetime isometries produce exact multisymplectomorphisms. 
So the multimomentum map on $J^1\pi$ is
\begin{equation*}
{\rm J}_\mathscr{L}(X_\zeta)=-i(X_\zeta)\Theta_\mathscr{L}=T\sqrt{-\text{det}g}\ G_{\mu\nu}G^{ba}x^\nu_b\xi^\mu\text{d}^1\sigma_a\in\df^1(J^1\pi)\ .
\end{equation*}
Furthermore, the exact multisymplectomorphisms of $J^{1*}\pi$ are generated by
\begin{equation*}
Y_\zeta=\mathscr{FL}_* X_\zeta=-\xi^\mu\frac{\partial}{\partial x^\mu}+\frac{1}{T}\sqrt{-\text{det}\Pi}\, G^{\mu\nu}\Pi_{ba}\, p^b_\mu\,\derpar{\xi_\mu}{x^\nu}\frac{\partial}{\partial p^a_\mu}
\in\vf(J^{1*}\pi)\ ,
\end{equation*}
and the corresponding multimomentum map is given by 
\begin{equation*}
{\rm J}_{\rm h}(Y_\zeta)=-i(Y_\zeta)\Theta_{\rm h}=-p^a_\mu\,\zeta^\mu\,\text{d}^1\sigma_a\in\df^1(J^{1*}\pi)\ .
\end{equation*}

\subsection{Yang--Mills theory}
\label{YangMills}

Yang--Mills theory with non-Abelian gauge group $SU(N)$ takes place on a $4$-dimensional Minkowski spacetime manifold $M$ 
endowed with the Minkowski metric $\eta_{\mu\nu}$ of signature $(-+++)$. 
The gauge connection is denoted as $A_\mu=A_\mu^at_a$, where $t_a$ are the generators (in some matrix representation) of the corresponding Lie algebra $\mathfrak{g}=\mathfrak{su}(N)$.
These generators satisfy $[t_b,t_c]=f^a_{bc}t_a$ $\big(a=1,...,\text{dim}G\big)$; where $f^a_{bc}$ are the structure constants.  
Then, $J^1\pi$ has local coordinates $(x^\mu,A^a_\nu, A^a_{\mu\nu})$ while on $J^{1*}\pi$ they are $(x^\mu,A^a_\nu, \pi_a^{\mu\nu})$.
The Lagrangian density is 
\begin{equation*}
\mathscr{L}=L\,\d^4x=-\frac{1}{4}F^{\mu\nu}_a F^a_{\mu\nu}\text{d}^4x\in\Omega^4(J^1\pi)\ ,
\end{equation*}
where $F^a_{\mu\nu}=A^a_{\mu\nu}-A^a_{\nu\mu}-f^a_{bc}A^b_\mu A^c_\nu$ give the non-Abelian field strength curvature tensor when pulled back by $j^1\phi$. 
The Yang--Mills Lagrangian is singular as the multi-Hessian is singular:
$$
\frac{\partial^2L}{\partial A^a_{\mu\nu}\partial A^b_{\rho\sigma}}=\delta_{ab}\left(\eta^{\mu\sigma}\eta^{\nu\rho}-\eta^{\mu\rho}\eta^{\nu\sigma}\right)\ .
$$
Using the notation in which rank-2 tensor components split up into their symmetric part $V_{(\mu\nu)}$ and antisymmetric part $V_{[\mu\nu]}$ as $V_{\mu\nu}=V_{(\mu\nu)}+V_{[\mu\nu]}$, it is evident that 
the relevant null vectors of the Hessian matrix above form the set of symmetric spacetime matrices $\left\{V_{\mu\nu}=V_{(\mu\nu)}\right\}$. 

The Lagrangian energy function $E_\mathscr{L}\in C^\infty(J^1\pi)$ and the Poincar\'e--Cartan forms on $J^1\pi$ are
\beann
E_\mathscr{L}&=&-F^{\mu\nu}_aA^a_{\mu\nu}+\frac{1}{4}F^{\mu\nu}_a F^a_{\mu\nu}\ . \\
\Theta_\mathscr{L}&=&-F^{\mu\nu}_a\text{d}A^a_\nu\wedge\text{d}^3x_\mu+\left(F^{\mu\nu}_a A^a_{\mu\nu}-\frac{1}{4}F^{\mu\nu}_a F_{\mu\nu}^a\right)\text{d}^4x\ ,\\
\Omega_\mathscr{L}&=&
-\left(\eta^{\mu\sigma}\eta^{\nu\rho}-\eta^{\mu\rho}\eta^{\nu\sigma}\right)\left(\text{d}A_{a\rho\sigma}+f_{abc}A^b_\sigma\text{d}A^c_\rho\right)\wedge\text{d}A^a_\nu\wedge\text{d}^3x_\mu -f^a_{bc}F^{\mu\nu}_aA^b_\mu\text{d}A^c_\nu\wedge\text{d}^4x\\
&\ &+\left(\eta^{\mu\sigma}\eta^{\nu\rho}-\eta^{\mu\rho}\eta^{\nu\sigma}\right)A_a^{\rho\sigma}\left(\text{d}A^a_{\mu\nu}\wedge\text{d}^4x+f^a_{bc}A^b_\mu\text{d}A^c_\nu\wedge\text{d}^4x\right)\ .
\eeann
The resulting field equations are the Euler--Lagrange equations,
$$
D^a_{\mu b}F^{\mu\nu}_a=0
$$
where $\displaystyle D^a_{\mu b}=\left(\derpar{}{x^\mu} -f^a_{cb}A^c_\mu\right)$.

The Legendre map $\mathscr{FL}:J^1\pi\rightarrow J^{1*}\pi$ is almost-regular and its image is the primary constraint submanifold $P_0\subset J^{1*}\pi$ defined by 
\begin{equation}
\label{YMp0}
\mathscr{FL}_*\pi^{\mu\nu}_a=-F^{\mu\nu}_a\ .
\end{equation}
It follows that, as $F^{\mu\nu}_a=F^{[\mu\nu]}_a$, the image of $\mathscr{FL}$ gives $\pi^{(\mu\nu)}=0$.
The Hamilton--Cartan forms are
\beann
\Theta^0_{\rm h}&=&\pi^{[\mu\nu]}_a\text{d}A^a_\nu\wedge\text{d}^3x_\mu-\left(\frac{1}{2}f^a_{bc}\pi^{[\mu\nu]}_aA^b_\mu A^c_\nu-\frac{1}{4}\pi^{[\mu\nu]}_a\pi_{[\mu\nu]}^a\right)\text{d}^4x\ , \\
\Omega^0_{\rm h}&=&-\text{d}\pi^{[\mu\nu]}_a\wedge\text{d}A^a_\nu\wedge\text{d}^3x_\mu+\left[\frac{1}{2}\left(f^a_{bc}A^b_\mu A^c_\nu-\pi^a_{[\mu\nu]}\right)\text{d}\pi^{[\mu\nu]}_a+f^a_{bc}\pi^{[\mu\nu]}_a A^b_\mu\text{d}A^c_\nu\right]\wedge\text{d}^4x\ .
\eeann
There are no Hamiltonian constraints other than the primary constraints \eqref{YMp0}{\color{red},} hence $P_f=P_0$. 
The (Hamilton)--De Donder--Weyl field equations for this theory are
$$
D^a_{\mu b}\pi^{[\mu\nu]}_a=0\quad , \quad
\derpar{A^a_\nu}{x^\mu} +\frac{1}{2}\left(\pi_{\mu\nu}^a-f^a_{bc}A^b_\mu A^c_\nu\right)=0\ .
$$

The $SU(N)$ gauge fields transform as
$$
\delta A_\mu=\derpar{\chi}{x^\mu}+[A_\mu,\chi]\equiv D_\mu\chi \quad ,\quad
\delta A^a_\mu=\delta^a_{\ b}\derpar{\chi^b}{x^\mu}-f^a_{cb}\chi^bA^c_\mu\equiv D^a_{\mu b}\chi^b\ ,
$$
The vector field which generates the $SU(N)$ transformations on $E$ is
\begin{equation*}
X^E_\chi=-D^a_{\mu b}\chi^b\frac{\partial}{\partial A^a_\nu}\in\mathfrak{X}(E)\ ,
\end{equation*}
the canonical lift of $X^E_\chi$ to $J^1\pi$ is given by
\begin{equation*}
X_\chi=j^1X^E_\chi=-D^a_{\mu b}\chi^b \frac{\partial}{\partial A^a_\mu}-\left(\derpar {D^a_{\nu b}\chi^b}{x^\mu}-f^a_{cb}A^c_{\mu\nu}\chi^b\right)\frac{\partial}{\partial A^a_{\mu\nu}}\in\mathfrak{X}(J^1\pi)\ ,
\end{equation*}
and the $\mathscr{FL}$-projection onto $P_0\subset J^{1*}\pi$, $Y_\chi=\mathscr{FL}_*X_\chi\in\mathfrak{X}(P_0)$, is
\begin{equation*}
Y_\chi=-D^a_{\mu b}\chi^b \frac{\partial}{\partial A^a_\mu}-\left[\derpar {D^a_{\nu b}\chi^b}{x^\mu}+\frac{1}{2}f^a_{cb}\chi^b\left(\pi^c_{[\mu\nu]}-f^c_{de}\chi^b A^d_\mu A^e_\nu\right)\right]\left(\delta^{\mu\beta}\delta^{\nu\alpha}-\delta^{\mu\alpha}\delta^{\nu\beta}\right)\derpar{}{\pi^{[\alpha\beta]}_a}\ .
\end{equation*}
The corresponding multimomentum maps are
\beann
{\rm J}_\mathscr{L}(X_\chi)&=&-i(X_\chi)\Theta_\mathscr{L}=-F^{\mu\nu}_aD^a_{\nu b}\chi^b\text{d}^3x_\mu \in\Omega^3(J^1\pi)\ ,\\
{\rm J}^0_{\rm h}(Y_\chi)&=&-i(Y_\chi)\Theta^0_{\rm h}=\pi^{[\mu\nu]}_aD^a_{\nu b}\chi^b\text{d}^3x_\mu\in\Omega^3(P_0)\ .
\eeann

Yang-Mills also exhibits spacetime isometries as a Noether symmetry. The infinitesimal version of this symmetry begins by lifting Killing vectors, in this case of the Minkowski metric, $\xi=-\xi^\mu(x)\derpar{}{x^\mu}\in\vf(M)$, to the configuration bundle,
$$
\xi_E=-\xi^\mu(x)\derpar{}{x^\mu}+A^a_\nu\derpar{\xi^\nu}{x^\mu}\derpar{}{A^a_\mu}\in\vf(E)\ .
$$
It follows that the canonical lift to the multivelocity phase space $J^1\pi$ is given as
$$
X_\xi=j^1\xi_E=-\xi^\mu(x)\derpar{}{x^\mu}+A^a_{\nu}\derpar{\xi^\nu}{x^\mu}\derpar{}{A^a_\mu}+\left(A^a_{\nu\mu}\derpar{\xi^\nu}{x^\rho}+A^a_{\rho\nu}\derpar{\xi^\nu}{x^\mu}\right)\derpar{}{A^a_{\rho\mu}}\ ,
$$
and the corresponding multimomentum map on $J^1\pi$ is 
$$
{\rm J}_\mathscr{L}(X_\xi)=-i(X_\xi)\Theta_\mathscr{L}=F^{\mu\nu}_a A^a_\rho\derpar{\xi^\rho}{x^\nu}\text{d}^3x_\mu+F^{\mu\nu}_a\xi^\rho \text{d}A^a_\nu\wedge\text{d}^2x_{\mu\rho}+\left(F^{\mu\nu}_aA^a_{\mu\nu}-\frac{1}{4}F^a_{\mu\nu}F^{\mu\nu}_a\right)\xi^\rho\text{d}^3x_\rho\ .
$$
The the $\mathscr{FL}$ projection of $X_\xi$ is
\beann
Y_\xi &=&\mathscr{FL}_*X_\xi=-\xi^\mu(x)\derpar{}{x^\mu} +A^a_{\nu}\derpar{\xi^\nu}{x^\mu}\derpar{}{A^a_\mu}
\\ 
&&+\eta^{\alpha\mu}\eta^{\beta\rho}\left[\left(\pi_{b[\mu\nu]}-f_{bcd}A^c_\mu A^d_\nu\right)\derpar{\xi^\nu}{x^\rho}+\left(\pi_{b[\nu\rho]}-f_{bcd}A^c_\nu A^d_\rho\right)\derpar{\xi^\nu}{x^\mu}\right]\derpar{}{\pi^{[\alpha\beta]}_b}\ ,
\eeann
and the resulting multimomentum map on $P_0\subset J^1\pi^*$ is 
$$
{\rm J}^0_h(Y_\xi)=-\pi^{[\mu\nu]}_a A^a_\rho\derpar{\xi^\rho}{x^\nu}\text{d}^3x_\mu-\pi^{[\mu\nu]}_a\xi^\rho \text{d}A^a_\nu\wedge\text{d}^2x_{\mu\rho}-\left(\pi^{[\mu\nu]}_aA^a_{\mu\nu}+\frac{1}{4}\pi^a_{[\mu\nu]}\pi^{[\mu\nu]}_a\right)\xi^\rho\text{d}^3x_\rho\ .
$$

\subsection{Chern--Simons theory}
\label{Ex3}

For the multisymplectic treatment of Chern-Simons theory with the Abelian gauge group $U(1)$ see \cite{GIMMSY}.
Consider Chern-Simons theory in $2+1$ dimensions with gauge group $G=SU(N)$, $\text{dim}\,G=N^2-1$. 
The gauge connection is denoted as $A_\mu=A_\mu^at_a$ where $t_a$ are the generators (in some matrix representation) of the corresponding Lie algebra $\mathfrak{g}=\mathfrak{su}(N)$. 
The generators satisfy $[t_b,t_c]=f^a_{bc}t_a$, $a=1,...,\text{dim}\,G$,
where $f^a_{bc}$ are structure constants.  
Then, $J^1\pi$ has local coordinates $(x^\mu,A^a_\nu, A^b_{\mu\nu})$. Now, let $A=A_\mu\text{d}x^\mu\in \Omega^1(J^1\pi)$ and $F=\frac{1}{2}F_{\mu\nu}\text{d}x^\mu\wedge\text{d}x^\nu\in \Omega^2(J^1\pi)$, with $\displaystyle F_{\mu\nu}\equiv 2A_{[\mu\nu]}+\frac{1}{2}[A_{\mu}, A_{\nu}]$; so the Chern-Simons Lagrangian density is
\begin{subequations}
\begin{align*}
\mathscr{L}&=L(x^\mu,A_\nu,A_{\mu\nu})\text{d}^3x=
\text{Tr}\left[F\wedge A -\frac{1}{3}A\wedge A\wedge A\right]=\epsilon^{\mu\nu\rho}\text{Tr}\left[A_{\mu\nu}A_\rho+\frac{2}{3}A_\mu A_\nu A_\rho\right]\text{d}^3x\\
&=\epsilon^{\mu\nu\rho}g_{ab}\left(-\frac{1}{4}F^a_{\mu\nu}A^b_\rho+\frac{1}{6}f^{b}_{cd}A^a_\mu A^c_\nu A^d_\rho\right)\text{d}^3x=-\frac{1}{2}\epsilon^{\mu\nu\rho}g_{ab}\left(A^a_{\mu\nu}A^b_\rho+\frac{2}{3}f^{b}_{cd}A^a_\mu A^c_\nu A^d_\rho\right)\text{d}^3x\ ,
\end{align*}
\end{subequations}
where the trace shown above is ad-invariant on the Lie algebra and can thereby be taken using the {\sl Cartan-Killing metric} 
$g_{ab}=-2\text{Tr}(t_at_b)$ on $\mathfrak{g}$.
This theory is singular as the Hessian is
\begin{equation*}
\frac{\partial^2L}{\partial A^a_{\mu\nu}\partial A^b_{\rho\sigma}}=0\ .
\end{equation*}
The Lagrangian energy and the Poincar\'e-Cartan forms are given by
\begin{subequations}
\begin{align*}
&E_{\mathscr{L}}\equiv \frac{\partial L}{\partial A^a_{\mu\nu}} A^a_{\mu\nu} - L=\frac{1}{3}\epsilon^{\mu\nu\rho}g_{ab}f^b_{cd}A^a_\mu A^c_\nu A^d_\rho\in C^\infty(J^1\pi) \ , \\
&\Theta_\mathscr{L}=-\epsilon^{\mu\nu\rho}g_{ab}\left[\frac{1}{2}A^b_\rho\text{d}A^a_\nu\wedge\text{d}^2x_\mu+\frac{1}{3}f^b_{cd}A^a_\mu A^c_\nu A^d_\rho\text{d}^3x\right]\ ,\\
&\Omega_\mathscr{L}=\epsilon^{\mu\nu\rho}g_{ab}\left[\frac{1}{2}\text{d}A^b_\rho\wedge\text{d}A^a_\nu\wedge\text{d}^2x_\mu + f^b_{cd}A^a_\mu A^c_\nu \text{d}A^d_\rho\wedge \text{d}^3x\right]\ .
\end{align*}
\end{subequations}
The Lagrangian constraint in this theory arises from imposing the {\sc sopde} condition on the multivector field solutions to the Lagrangian field equations, giving the following constraint
\begin{equation*}
\epsilon^{\mu\nu\rho}\left(g_{ab}A^b_{\mu\nu}+g_{db}f^b_{ca}A^d_\mu A^c_\nu\right)=0\ ,
\end{equation*}
which defines the submanifold ${\cal S}_f\subset J^1\pi$.
The tangency condition on this submanifold gives no new constraints.

The multimomentum phase space $J^{1*}\pi$ has local coordinates $(x^\mu, A^a_\nu,\pi_a^{\mu\nu})$ and the Legendre map 
$\mathscr{FL}:J^1\pi\rightarrow J^1\pi^*:(x^\mu, A^a_\nu,A^a_{\mu\nu})\mapsto (x^\mu, A^a_\nu,\pi_a^{\mu\nu})$ satisfies
\begin{equation*}
\mathscr{FL}^*\pi_a^{\mu\nu}=\frac{\partial \mathscr{L}}{\partial A^a_{\mu\nu}}=-\frac{1}{2}\epsilon^{\mu\nu\rho}g_{ab}A^b_\rho\ ,
\end{equation*}
which means that the image of $\mathscr{FL}$ is the submanifold $P_0\subset J^{1*}\pi$ defined by the constraints
\begin{equation*}
\pi_a^{\mu\nu}+\frac{1}{2}\epsilon^{\mu\nu\rho}g_{ab}A^b_\rho=0 \ .
\end{equation*}
Thus, local coordinates on $P_0$ are $(x^\mu, A^a_\nu)$.
The De Donder--Weyl Hamiltonian function is
\begin{equation*}
H_0(x^\mu,A_\nu)=\frac{1}{3}\epsilon^{\mu\nu\rho}g_{ab}f^b_{cd}A^a_\mu A^c_\nu A^d_\rho\in C^\infty(P_0)\ ,
\end{equation*}
 and the Hamilton-Cartan forms on $P_0$ are
\begin{subequations}
\begin{align*}
&\Theta^0_{\rm h}=-\epsilon^{\mu\nu\rho}g_{ab}\left[\frac{1}{2}A^b_\rho\text{d}A^a_\nu\wedge\text{d}^2x_\mu+\frac{1}{3}f^b_{cd}A^a_\mu A^c_\nu .x;p03A^d_\rho\text{d}^3x\right]\ ,\\
&\Omega^0_{\rm h}=\epsilon^{\mu\nu\rho}g_{ab}\left[\frac{1}{2}\text{d}A^b_\rho\wedge\text{d}A^a_\nu\wedge\text{d}^2x_\mu + f^b_{cd}A^a_\mu A^c_\nu \text{d}A^d_\rho\wedge \text{d}^3x\right]\ .
\end{align*}
\end{subequations}
There are no further Hamiltonian constraints other than the primary constraints
and $\mathscr{FL}({\cal S}_f)=P_0$. 
The field equations in both the Langrangian and Hamiltonian settings are
\begin{equation*}
\epsilon^{\mu\nu\nu}g_{ab}\left(\partial_\mu A^b_{\nu}+f^b_{dc}A^d_\mu A^c_\nu\right)=0\ \Rightarrow\ F^b_{\mu\nu}=0\ \Rightarrow\ F_{\mu\nu}=0\ .
\end{equation*}

The gauge symmetry group $SU(N)$ of the Chern-Simons theory produces infinitesimal transformations
$$
\delta A_\mu=D_\mu\chi=\partial_\mu\chi+[A_\mu,\chi]\quad ,\quad
\delta A^a_\mu=D^a_{\mu b}\chi^b=\delta^a_{\ b}\partial_\mu \chi^b-f^a_{cb}\chi^bA^c_\mu\ ,
$$
where $\chi=\chi^at_a\in\Cinfty(M)$. 
The vector field which generates the $SU(N)$ transformations on $E$ is
\begin{equation*}
X^E_\chi=-D^a_{\mu b}\chi^b\frac{\partial}{\partial A^a_\nu}\in\mathfrak{X}(E)\ ,
\end{equation*}
whose canonical lift to $J^1\pi$ is
\begin{equation*}
X_\chi=j^1X^E_\chi=-D^a_{\mu b}\chi^b \frac{\partial}{\partial A^a_\mu}-\left(\derpar {D^a_{\nu b}\chi^b}{x^\mu}-f^a_{cb}A^c_{\mu\nu}\chi^b\right)\frac{\partial}{\partial A^a_{\mu\nu}}\in\mathfrak{X}(J^1\pi)\ .
\end{equation*}
and its $\mathscr{FL}$-projection onto $P_0\subset J^{1*}\pi$ is
\begin{equation*}
Y_\chi=\mathscr{FL}_*X_\chi=-D^a_{\mu b}\chi^b \frac{\partial}{\partial A^a_\mu}\in\mathfrak{X}(P_0)\ .
\end{equation*}
Finally, the corresponding multimomentum maps are
\beann
{\rm J}_\mathscr{L}(X_\chi)&=&-i(X_\chi)\Theta_\mathscr{L}=-\frac{1}{2}\epsilon^{\mu\nu\rho}g_{ab}A^b_\rho D^a_{\nu b}\chi^b\text{d}^2x_\mu
\in\Omega^2(J^1\pi)\ ,\\
{\rm J}^0_{\rm h}(Y_\chi)&=&-i(Y_\chi)\Theta^0_{h}=-\frac{1}{2}\epsilon^{\mu\nu\rho}g_{ab}A^b_\rho D^a_{\nu b}\chi^b\text{d}^2x_\mu
\in\Omega^2(J^1\pi)
\in\Omega^2(P_0)\ .
\eeann

Exactly as in Yang--Mills theory, spacetime diffeomorphisms which are isometries are a Noether symmetry generated by the Killing vector field $\displaystyle \xi=-\xi^\mu(x)\frac{\partial}{\partial x^\mu}\in\mathfrak{X}(M)$.
The canonical lift to the configuration bundle $E$ is
$$
\displaystyle \xi_E=-\xi^\mu(x)\frac{\partial}{\partial x^\mu}+A^a_{\nu}\derpar{\xi^\nu}{x^\mu}\frac{\partial}{\partial A^a_\nu}\in\mathfrak{X}(E)\ ,
$$ 
which can be lifted again to $J^1\pi$, giving 
$$
X_\xi=j^1\xi_E=-\xi^\mu(x)\derpar{}{x^\mu}+A^a_{\nu}\derpar{\xi^\nu}{x^\mu}\derpar{}{A^a_\mu}+\left(A^a_{\nu\mu}\derpar{\xi^\nu}{x^\rho}+A^a_{\rho\nu}\derpar{\xi^\nu}{x^\mu}\right)\derpar{}{A^a_{\rho\mu}}\in\mathfrak{X}(J^1\pi)\ ,
$$
as before.
The corresponding multimomentum map ${\rm J}_\mathscr{L}(X_\xi)\in\Omega^2(J^1\pi)$ is 
\beann
{\rm J}_\mathscr{L}(X_\xi)&=&-\inn(X_\xi)\Theta_\mathscr{L} \\
&=&\epsilon^{\mu\nu\rho}g_{ab}\left[\frac{1}{2}A^b_\rho\left(A^a_\sigma\derpar{\xi^\sigma}{x^\nu}\text{d}^2x_\mu + \xi^\sigma\text{d}A^a_\nu\wedge\text{d}^1x_{\mu\sigma}\right)-\frac{1}{3}f^b_{cd}\xi^\sigma A^a_\mu A^c_\nu A^d_\rho\text{d}^2x_\sigma\right]
\ .
\eeann
Furthermore, the projection of $X_\xi\in\vf(J^1\pi)$ to $P_0\subset J^1\pi^*$ is
$$
Y_\xi=\mathscr{FL}_* X_\xi=-\xi^\mu(x)\derpar{}{x^\mu}+A^a_{\nu}\derpar{\xi^\nu}{x^\mu}\derpar{}{A^a_\mu}\in\vf(P_0)\ .
$$
The resulting multimomentum map ${\rm J}^0_{\rm h}(Y_\xi)\in\Omega^2(P_0)$ is
\beann
{\rm J}^0_{\rm h}(Y_\xi)&=&-i(Y_\xi)\Theta^0_{\rm h} \\
&=&\epsilon^{\mu\nu\rho}g_{ab}\left[\frac{1}{2}A^b_\rho\left(A^a_\sigma\derpar{\xi^\sigma}{x^\nu}\text{d}^2x_\mu + \xi^\sigma\text{d}A^a_\nu\wedge\text{d}^1x_{\mu\sigma}\right)-\frac{1}{3}f^b_{cd}\xi^\sigma A^a_\mu A^c_\nu A^d_\rho\text{d}^2x_\sigma\right]\ .
\eeann

\subsection{Electric Carrollian scalar field theory}
\label{electriccarroll}

Consider two scalar fields $\phi(x)$ and $\pi(x)$ which are given as the local sections $\phi:M\rightarrow E: x^\mu\mapsto (x^\mu,\phi(x), \pi(x))$ of a configuration bundle $\pi:E\rightarrow M: (\phi, \pi) \mapsto x^\mu$ 
where $M$ is an $m$-dimensional spacetime manifold with Minkowski metric $\eta_{\mu\nu}$. Now, take the {\sl electric Carrollian contraction} of the following Lagrangian,
\label{Ex4}
\begin{equation}
\label{scalarLag}
L=\pi\phi_0-\frac{1}{2}\pi^2-\frac{1}{2}\phi_i\phi^i \in C^\infty (J^1\pi)\ ,
\end{equation}
by making the field redefinition $\phi(x)\rightarrow c\phi(x)$, $\pi(x)\rightarrow \frac{1}{c}\pi(x)$, and taking the limit $c\rightarrow 0$ for the speed of light. 
It follows that the Minkowski metric becomes degenerate:
\begin{equation}
\label{carrollian metric}
\text{d}s^2=\eta_{\mu\nu}\text{d}x^\mu\otimes\text{d}x^\nu=-c^2\text{d}x^0\otimes\text{d}x^0+\delta_{ij}\text{d}x^i\otimes\text{d}x^j\ \longrightarrow\  \delta_{ij}\text{d}x^i\otimes\text{d}x^j\ .
\end{equation}
The Lagrangian function $L\in C^\infty (J^1\pi)$ which is obtained from the electric Carrollian contraction of \eqref{scalarLag} is
$$
L=\pi\phi_0-\frac{1}{2}\pi^2\ ,
$$
which is the electric Carrollian scalar field Lagrangian studied in this section. 
This Lagrangian is singular since
$\displaystyle\frac{\partial^2L}{\partial\phi_\mu\partial\phi_\nu}=\displaystyle\frac{\partial^2L}{\partial\pi_\mu\partial\pi_\nu}=\displaystyle\frac{\partial^2L}{\partial\pi_\mu\partial\phi_\nu}=0$.
The Lagrangian energy function and the Poincar\'e--Cartan forms are
\beann
E_\mathscr{L}&=&\frac{1}{2}\pi^2\in C^\infty(J^1\pi) \ , \\
\Theta_\mathscr{L}&=&\pi\text{d}\phi\wedge\text{d}^{m-1}x_0-\frac{1}{2}\pi^2\text{d}^mx\in \Omega^m(J^1\pi) \ , \\
\Omega_\mathscr{L}&=&\text{d}\phi\wedge\text{d}\pi\wedge\text{d}^{n-1}x_0+\pi\text{d}\pi\wedge\text{d}^mx\in \Omega^{m+1(}J^1\pi)\ .
\eeann
Again, the compatibility of the Lagrangian field equations produce two {\sc sopde} constraints
$$
\pi_0=0 \quad ,\quad \phi_0-\pi=0 \ ,
$$
which define the constraint submanifold ${\cal S}_f\subset J^1\pi$,
and the tangency condition on this submanifold gives no new constraints.

The Legendre map is
\begin{equation*}
\mathscr{FL}^*p_\phi^0\equiv \frac{\partial L}{\partial\phi_0}=\pi
\quad ,\quad
\mathscr{FL}^*p_\phi^i=\frac{\partial L}{\partial\phi_i}=0
\quad ,\quad
\mathscr{FL}^*p^\mu_\pi=\frac{\partial L}{\partial \pi_\mu}=0 \ ,
\end{equation*}
and give the primary constraints
$$
p_\phi^0-\pi=0 \quad , \quad p^i_\phi=0 \quad, \quad p^\mu_\pi=0 \ ,
$$
which define the submanifold $P_0\subset J^1\pi^*$ with local coordinates $(x^\mu,\phi,\pi)$. 
The Lagrangian function $L$ is almost-regular and the De Donder--Weyl Hamiltonian is 
\begin{equation*}
H_0=\frac{1}{2}\pi^2\in\Cinfty(P_0) \ ,
\end{equation*}
and the Hamilton-Cartan forms are 
\beann
\Theta_{\rm h}^0&=&\pi\text{d}\phi\wedge\text{d}^{m-1}x_0-\frac{1}{2}\pi^2\text{d}^mx\in\df^m(P_0) \ , \\
\Omega_{\rm h}^0&=&\text{d}\phi\wedge\text{d}\pi\wedge\text{d}^{m-1}x_0+\pi\text{d}\pi\wedge\text{d}^mx \ .
\eeann

The spacetime symmetries are the Carroll transformations \cite{CarrollScalar}
\beq
\label{carroll1}
x'^0=x^0+b_kx^k \quad, \quad x'^i=x^i+\epsilon^i_{\ j}x^j
\quad \Rightarrow \quad x'^{\mu}=C^\mu_{\ \nu}x^\nu \ ,
\eeq
where $C^0_{\ 0}=1$, $C^0_{\ k}=b_k$, $C^k_{\ 0}=0$, $(C^i_{\ j})\in O(d)$.
These transformations are represented infinitesimally by
\begin{equation}
\label{carroll2}
C^\mu_{\ \nu}= \delta^\mu_{\ \nu}+\epsilon^\mu_{\ \nu}\ ,
\end{equation}
so that now $b_k=\epsilon^0_{\ k}$ is infinitesimal. 
Up to linear terms in $\phi'(x')=\phi(x)$ and $\pi'(x')=\pi(x)$, the infinitesimal transformations of the fields are
\begin{align*}
\delta \phi(x)&\equiv \phi'(x)-\phi(x)=-\epsilon_{\ \beta}^{\alpha}x^{\beta}\derpar{\phi}{x^\alpha}=- b_kx^k\derpar{\phi}{x^0}-\epsilon^i_{\ j}x^j\derpar{\phi}{x^i} \ , \\
\delta \pi(x)&\equiv \pi'(x)-\pi(x)=-\epsilon_{\ \beta}^{\alpha}x^\beta\derpar{\pi}{x^\alpha}= -b_kx^k\derpar{\pi}{x^0}-\epsilon^i_{\ j}x^j\derpar{\pi}{x^i} \ .
\end{align*}
Furthermore, 
\begin{equation*}
\phi'_\mu(x')=\frac{\partial x^\nu}{\partial x'^\mu}\phi_\nu(x)\ \Longleftrightarrow\ \phi'_\mu(C\cdot x) =(C^{-1})_{\ \mu}^{\nu}\phi_\nu(x)\ ,
\end{equation*}
where $C\equiv(C_{\ \nu}^\mu)$ and $C\cdot x\equiv C_{\ \nu}^\mu x^\nu$; 
hence,
\begin{align*}
\delta\phi_\mu(x)&\equiv \phi'_\mu(x)-\phi_\mu(x)= -\epsilon^\nu_{\ \mu}\phi_\nu-\epsilon_{\ \beta}^{\alpha}x^{\beta}\derpar{\phi_\mu}{x^\alpha} \ ,\\ 
\delta \pi_\mu(x)&\equiv \pi'_\mu(x)-\pi_\mu=-\epsilon^\nu_{\ \mu}\pi_\nu -\epsilon_{\ \beta}^{\alpha}x^{\beta}\derpar{\pi_\mu}{x^\alpha} \ .
\end{align*}
Component-wise, 
\begin{align*}
\delta\phi_0(x)&=-\epsilon_{\ \beta}^{\alpha}x^{\beta}\derpar{\phi_0}{x^\alpha}=-b_kx^k\derpar{\phi_0}{x^0}-\epsilon^k_{\ j}x^j\derpar{\phi_0}{x^k} \ ,\\ 
\delta\phi_i(x)&=-\epsilon^\nu_{\ i}\phi_\nu-\epsilon_{\ \beta}^{\alpha}x^{\beta}\derpar{\phi_i}{x^\alpha}=-b_i\phi_0-\epsilon^k_{\ i}\phi_k- b_kx^k\derpar{\phi_i}{x^0}-\epsilon^k_{\ j}x^j\derpar{\phi_i}{x^k} \ ,
\end{align*}
and similarly for the components of $\delta \pi_\mu(x)$. Under the transformations presented above, 
the Lagrangian transforms as $\delta\mathscr{L}(x)=-\partial_\alpha\left(\epsilon^\alpha_{\ \beta}x^\beta\mathscr{L}(x)\right)$. 
These transformations are produced by the Lie derivatives of the local sections $\phi:M\rightarrow E$ 
and $j^1\phi:M\rightarrow J^1\pi$ by a vector field $\xi=-\xi^\mu(x)\derpar{}{x^\mu}\in\vf(M)$ 
which generates the Carroll transformations; where now $\xi^\mu(x)=\epsilon^\mu_{\ \nu}x^\nu$ is given by (\ref{carroll1}) and (\ref{carroll2}). 
The vector field which generates the Carroll transformations on the configuration bundle $E$ is
\begin{equation*}
\xi_E=-\epsilon^\mu_{\ \nu}x^\nu\frac{\partial}{\partial x^\mu}\in \vf(E)\ ,
\end{equation*}
while the canonical lift of $\xi_E$ to $J^1\pi$ is given by
\begin{align*}
\begin{split}
X&=-\epsilon^\mu_{\ \nu}x^\nu\frac{\partial}{\partial x^\mu} +\epsilon^\nu_{\ \mu}\phi_\nu\frac{\partial}{\partial\phi_\mu}+\epsilon^\nu_{\ \mu}\pi_\nu\frac{\partial}{\partial \pi_\mu} \\
&=-b_ix^i\frac{\partial}{\partial x^0}-\epsilon^i_{\ j}x^j\frac{\partial}{\partial x^i}+\left(b_i\phi_0+\epsilon_{\ i}^{ j}\phi_j\right)\frac{\partial}{\partial \phi_i}+\left(b_i\pi_0+\epsilon_{\ i}^{ j}\pi_j\right)\frac{\partial}{\partial \pi_i}\in\vf(J^1\pi) \ .
\end{split}
\end{align*}
Evidently $X$ is tangent to the constraint submanifold ${\cal S}_f$ since $\Lie(X)(\pi-\phi_0)=0$ and $\Lie(X)\pi_0=0$.
Finally, since $\Lie(X)\mathscr{L}=0 \Rightarrow \Lie(X)\Theta_\mathscr{L}=0$, 
the momentum map ${\rm J}_\mathscr{L}(X)$ corresponding to the Carroll spacetime transformations in the Lagrangian setting is given by
\begin{equation*}
{\rm J}_\mathscr{L}(X)=-i(X)\Theta_\mathscr{L}=-\epsilon^i_{\ j}x^j\pi\text{d}\phi\wedge\text{d}^{m-2}x_{0i}-\frac{1}{2}\pi^2\left(b_ix^i\text{d}^{m-1}x_0+\epsilon^i_{\ j}x^j\text{d}^{m-1}x_i\right)\in\df^{m-1}(J^1\pi) \ .
\end{equation*}

In the Hamiltonian setting, which takes place on $P_0\subset J^1\pi^*$, the momentum map ${\rm J}^0_{\rm h}(Y)\in\df^{m-1}(P_0)$ is constructed using the vector field
\begin{equation*}
Y=\mathscr{FL}_{0*}X=-\epsilon^\mu_{\ \nu}x^\nu\derpar{}{x^\mu}=-b_ix^i\frac{\partial}{\partial x^0}-\epsilon^i_{\ j}x^j\frac{\partial}{\partial x^i}\in\vf(P_0)\ .
\end{equation*}
As in the Lagrangian setting, $\Lie(Y)\Theta^0_{\rm h}=0$ 
and ${\rm J}^0_{\rm h}(Y)$ is given by
\begin{equation*}
{\rm J}^0_{\rm h}(Y)=-i(Y)\Theta_{\rm h}^0=-\epsilon^i_{\ j}x^j\pi\,\text{d}\phi\wedge\text{d}^{m-2}x_{0i}-\frac{1}{2}\pi^2\left(b_ix^i\text{d}^{m-1}x_0+\epsilon^i_{\ j}x^j\text{d}^{m-1}x_i\right)\ .
\end{equation*}

\subsection{Magnetic Carrollian scalar field theory}
\label{Ex5}

The {\sl magnetic Carrollian contraction} \cite{CarrollScalar} of the canonical Klein--Gordon Lagrangian is performed by reinserting the factors of $c$ (speed of light) 
into the Lagrangian (\ref{scalarLag}) and taking the limit $c\rightarrow 0$. The Minkowski metric (\ref{carrollian metric}) becomes degenerate as in the electric Carrollian scalar field theory. 
The Lagrangian function $L\in C^\infty (J^1\pi)$ obtained from taking the magnetic Carrollian contraction of the canonical Klein--Gordon Lagrangian (\ref{scalarLag}) is
\begin{equation*}
L=\pi\phi_0-\frac{1}{2}\phi_i\phi^i \ .
\end{equation*}
This Lagrangian is singular since the components of the Hessian
matrix (with respect to the multivelocities) are
\begin{equation*}
\frac{\partial^2L}{\partial\phi_0\partial\phi_0}=0\quad , \quad 
\frac{\partial^2L}{\partial\phi_0\partial\phi_i}=0\quad , \quad 
\frac{\partial^2L}{\partial\phi_i\partial\phi_j}=-\delta^{ij}\quad , \quad 
\frac{\partial^2L}{\partial\pi_\mu\partial\pi_\nu}=0\quad , \quad 
\frac{\partial^2L}{\partial\pi_\mu\partial\phi_\nu}=0\ . 
\end{equation*}

The Lagrangian energy function and the Poincar\'e--Cartan forms are now
\beann
E_\mathscr{L}&=&-\frac{1}{2}\phi_i\phi^i\in C^\infty(J^1\pi)  \ ,
\\
\Theta_\mathscr{L}&=&\pi\text{d}\phi\wedge\text{d}^{m-1}x_0-\phi^i\text{d}\phi\wedge\text{d}^{m-1}x_i+\frac{1}{2}\phi_i\phi^i\text{d}^mx \in \Omega^m(J^1\pi)  \ ,
\\
\Omega_\mathscr{L}&=&\text{d}\phi\wedge\text{d}\pi\wedge\text{d}^{m-1}x_0+\text{d}\phi^i\wedge\text{d}\phi\wedge\text{d}^{m-1}x_i+\phi^i\text{d}\phi_i\wedge\text{d}^mx
\in \Omega^{m+1}(J^1\pi) ,
\eeann
and $\Omega_\mathscr{L}$ is a premultisymplectic form.
Imposing the {\sc sopde} condition,
\begin{equation*}
C_i^i-\pi_0=0 \quad ,\quad \phi_0=0 \ .
\end{equation*}
The first equation gives a relation for the coefficients $C_i^i$,
and the second one is a {\sc sopde}-constraint 
which defines the constraint submanifold $S_1\hookrightarrow J^1\pi$
and no more constraints arise from the constraint algorithm.

The Legendre map $\mathscr{FL}:J^1\pi\rightarrow J^1\pi^*$ gives the multimomenta
\begin{equation*}
\mathscr{FL}^*p_\phi^0= \frac{\partial\mathscr{L}}{\partial\phi_0}=\pi
\quad , \quad
\mathscr{FL}^*p_\phi^i=\frac{\partial\mathscr{L}}{\partial\phi_i}=-\phi^i
\quad , \quad
\mathscr{FL}^*p^\mu_\pi=\frac{\partial\mathscr{L}}{\partial \pi_\mu}=0 \ .
\end{equation*}
It follows that $p_\phi^0-\pi=0$ and $p_\pi^\mu=0$ are primary constraints which define the primary constraint submanifold $P_0\hookrightarrow J^1\pi^*$
with local coordinates $(x^\mu,\phi,\pi,p_\phi^i)$ and the Lagrangian is again almost-regular. 
The De Donder--Weyl Hamiltonian $H_0\in C^\infty(P_0)$ is given by
\begin{equation*}
H_0=-\frac{1}{2}p_{\phi i}p_\phi^i \ ,
\end{equation*}
and the Hamilton--Cartan forms on $P_0$ are 
\beann
\Theta_{\rm h}^0&=&\pi\text{d}\phi\wedge\text{d}^{m-1}x_0+p_\phi^i\text{d}\phi\wedge\text{d}^{m-1}x_i+\frac{1}{2}p_{\phi i}p_\phi^i\text{d}^mx\in \df^m(P_0) \ , \\
\Omega_{\rm h}^0&=&\text{d}\phi\wedge\text{d}\pi\wedge\text{d}^{m-1}x_0+\text{d}\phi\wedge\text{d}p_\phi^i\wedge\text{d}^{m-1}x_i-p_{\phi i}\text{d}p_\phi^i\wedge\text{d}^mx\in\df^{m+1}(P_0)\,.
\eeann

The Carroll transformations which are symmetries for the magnetic scalar field theory are written slightly differently than for the electric scalar field and are
\begin{align*}
\delta \phi(x)&=-\epsilon_{\ \beta}^{\alpha}x^{\beta}\derpar{\phi}{x^\alpha}=- b_kx^k\derpar{\phi}{x^0}-\epsilon^i_{\ j}x^j\derpar{\phi}{x^i} \ , \\
\delta \pi(x)&=-\epsilon_{\ \beta}^{\alpha}x^\beta\derpar{\pi}{x^\alpha}-b_i\phi^i= -b_kx^k\derpar{\pi}{x^0}-\epsilon^i_{\ j}x^j\derpar{\pi}{x^i}-b_i\phi^i \ , \\
\delta\phi_0(x)&=-\epsilon_{\ \beta}^{\alpha}x^{\beta}\derpar{\phi_0}{x^\alpha}=-b_kx^k\derpar{\phi_0}{x^0}-\epsilon^k_{\ j}x^j\derpar{\phi_0}{x^k} \ ,\\ 
\delta\phi_i(x)&=-\epsilon^\nu_{\ i}\phi_\nu-\epsilon_{\ \beta}^{\alpha}x^{\beta}\derpar{\phi_i}{x^\alpha}=- b_i\phi_0-\epsilon^k_{\ i}\phi_k- b_kx^k\derpar{\phi_i}{x^0}-\epsilon^k_{\ j}x^j\derpar{\phi_i}{x^k} \ .
\end{align*}
Under these transformations the Lagrangian transforms as $\displaystyle\delta L(x)=-\derpar{}{x^\alpha}\Big(\epsilon^\alpha_{\ \beta}x^\beta L(x)\Big)$. 
Notice that the field $\pi(x)$ no longer transforms as a Carrollian scalar; instead, $\pi(x)$ transforms under a Carroll spacetime transformation $-\epsilon_{\ \beta}^{\alpha}x^\beta\derpar{\pi}{x^\alpha}$ plus a term $-b_i\phi^i(x)$ which cannot be represented on $E$. 
The field transformations behave geometrically as the Lie derivatives of the local sections $j^1\phi:M\rightarrow J^1\pi$.
The vector field generating the Carroll transformations on $J^1\pi$ is
\begin{align*}
\begin{split}
X&=-\epsilon^\mu_{\ \nu}x^\nu\frac{\partial}{\partial x^\mu} + b_i\phi^i\frac{\partial}{\partial \pi}+\epsilon^\nu_{\ \mu}\phi_\nu\frac{\partial}{\partial\phi_\mu}+\epsilon^\nu_{\ \mu}\pi_\nu\frac{\partial}{\partial \pi_\mu} \\
&=-b_ix^i\frac{\partial}{\partial x^0}-\epsilon^i_{\ j}x^j\frac{\partial}{\partial x^i}+ b_i\phi^i\frac{\partial}{\partial \pi}+\left(b_i\phi_0+\epsilon_{\ i}^{ j}\phi_j\right)\frac{\partial}{\partial \phi_i} +\epsilon^\nu_{\ \mu}\pi_\nu\frac{\partial}{\partial \pi_\mu}\in\vf(J^1\pi)\ ,
\end{split}
\end{align*}
which is tangent to the constraint submanifold $S_1\subset J^1\pi$, given by the constraint $\phi_0=0$,
since $\Lie(X)\phi_0=0$. 
Furthermore, although the vector field $X$ leaves the Lagrangian density invariant ($\Lie(X)\mathscr{L}=0$), it does not produce an exact Noether symmetry as 
$$
\Lie(X)\Theta_\mathscr{L}=b^i\phi_0(\phi_i\text{d}^mx-\text{d}\phi\wedge\text{d}^{m-1}x_i) \ .
$$
Moreover, the vector field $X$ as written above on all of $J^1\pi$ is not projectable onto $P_0\subset J^1\pi^*$. Instead, $X$ on $S_1$ given by
\begin{equation}
\left.X\right\vert_{S_1}=-b_ix^i\frac{\partial}{\partial x^0}-\epsilon^i_{\ j}x^j\frac{\partial}{\partial x^i}+ b_i\phi^i\frac{\partial}{\partial \pi}+\epsilon_{\ i}^{ j}\phi_j\frac{\partial}{\partial \phi_i}+\epsilon^\nu_{\ \mu}\pi_\nu\frac{\partial}{\partial \pi_\mu}\ ,
\label{XS1}
\end{equation}
is $\mathscr{ FL}_0$-projectable onto $P_0$, giving
\begin{equation*}
Y=\mathscr{FL}_{0*}X\vert_{S_1}
=-b_ix^i\frac{\partial}{\partial x^0}-\epsilon^i_{\ j}x^j\frac{\partial}{\partial x^i}-b_ip_\phi^i\frac{\partial}{\partial \pi}+\epsilon^{ i}_{\ j}p_\phi^j\frac{\partial}{\partial p_\phi^i}\in\mathfrak{X}(P_0) \ .
\end{equation*}
Now, letting $\widetilde{X}$ be the local extension of $X\vert_{S_1}$ to $J^1\pi$
whose coordinate expression is given by \eqref{XS1};
then, it follows that
$\Lie(\widetilde{X})\Theta_\mathscr{L}=0$ on $J^1\pi$.
It is thereby possible to define the momentum maps, giving
\beann
{\rm J}_\mathscr{L}(\widetilde{X})=-i(\widetilde{X})\Theta_\mathscr{L}&=&
-\left(\epsilon^i_{\ j}x^j\pi+b_jx^j\phi^i\right)\text{d}\phi\wedge\text{d}^{m-2}x_{0i}+\phi^i\epsilon^j_{\ k}x^k\text{d}\phi\wedge\text{d}^{m-2}x_{ij}\\
& &+\frac{1}{2}\phi_i\phi^i\left(b_kx^k\text{d}^{m-1}x_0+\epsilon^k_{\ j}x^j\text{d}^{m-1}x_k\right)\in\df^{m-1}(J^1\pi) \ ,
\\
{\rm J}^0_{\rm h}(Y)=-i(Y)\Theta^0_{\rm h}&=&-\left(\epsilon^i_{\ j}x^j\pi-b_jx^jp_\phi^i\right)\text{d}\phi\wedge\text{d}^{m-2}x_{0i}-p_\phi^i\epsilon^j_{\ k}x^k\text{d}\phi\wedge\text{d}^{m-2}x_{ij}\\
& &+\frac{1}{2}p_{\phi i}p_\phi^i\left(b_kx^k\text{d}^{m-1}x_0+\epsilon^k_{\ j}x^j\text{d}^{m-1}x_k\right)\in\df^{m-1}(P_0) \ .
\eeann

\section{Conclusions}

One of the primary objectives achieved in this work is the precise connection of how classical field theories are studied in the standard physics literature and how they are treated in the more modern works of (pre)multisymplectic geometry.
This work provides a general overview of Cartan (Noether) symmetries present in (pre)multisymplectic geometry and how they are used to describe Noether symmetries in classical field theories both in the Lagrangian and DeDonder--Weyl Hamiltonian formalisms.
In the literature of differential geometry, the treatment of Cartan symmetries is well known \cite{art:deLeon_etal2004,art:GPR-2016,GIMMSY,RWZ-2016}. 
However, subtleties arise in the treatment of Noether symmetries of field theories whose Lagrangians are singular; 
the constraints of such singular field theories are obtained from the premultisymplectic forms of the corresponding multiphase spaces  \cite{GGR-2022}.
In particular, the analysis of Noether symmetries in the presence of the constraints is described in this paper. 

More specifically, the Noether symmetries of classical field theories developed on the relevant (pre)multisymplectic multi-phase spaces are encoded by their conserved quantities, called multimomentum maps (as described by Noether's theorem). 
The construction of the relevant multimomentum maps is carried out explicitly for various singular field theories that are important in theoretical physics. 
These multimomentum maps give the standard Noether currents of classical field theories when pulled back by local sections which map from the base space of the bundle structure to the multi-phase spaces. 
Furthermore, the multimomentum maps associated with gauge symmetries are essential for carrying out a desired symmetry reduction procedure (or gauge fixing), e.g. \cite{EMR-18}; the formal development of such symmetry reduction procedures in premultisymplectic field theories continues to be an open area of research (see, for instance, \cite{Bl-21,Bl-22,CCIMS-2020}).

\appendix

\section{Multivector fields on manifolds and fiber bundles}
\label{append}

(See \cite{Ca96a,CIL99,art:Echeverria_Munoz_Roman98}).
Let $\mathscr{M}$ be an $N$-dimensional differentiable manifold.
The {\sl \textbf{$m$-multivector fields}} in $\mathscr{M}$ ($m\leq N$)
are the sections of the $m$-multitangent bundle
$\displaystyle\bigwedge^m\Tan\mathscr{M}:=\overbrace{\Tan\mathscr{M}\wedge\ldots\wedge\Tan\mathscr{M}}^m$; that is,
the skew-symmetric contravariant tensor fields of order $m$ in $\mathscr{M}$;
the set of which is denoted $\vf^m (\mathscr{M})$.
Then, if $\mathbf{X}\in\mathfrak{X}^m(\mathscr{M})$,
for every point $\bar y\in \mathscr{M}$,
there is an open neighbourhood $U\subset \mathscr{M}$ and
$X_1,\ldots ,X_r\in\mathfrak{X} (U)$ such that,
for $m \leqslant r\leqslant{\rm dim}\,\mathscr{M}$,
$$
\mathbf{X}\vert_{U}=\sum_{1\leq i_1<\ldots <i_m\leq r} f^{i_1\ldots i_m}X_{i_1}\wedge\ldots\wedge X_{i_m} \, ,
$$
with $f^{i_1\ldots i_m} \in C^\infty (U)$.
In particular, $\mathbf{X}\in\vf^m(\mathscr{M})$ is said to be a
{\sl \textbf{locally decomposable multivector field}} if
there exist $X_1,\ldots ,X_m\in\vf (U)$ such that $\mathbf{X}\vert_U=X_1\wedge\ldots\wedge X_m$.
The locally decomposable $m$-multivector fields are locally associated with $m$-dimensional
distributions $D\subset\Tan \mathscr{M}$, and this splits
$\vf^m (\mathscr{M})$ into {\sl equivalence classes} $\{ {\bf X}\}\subset\vf^m(\mathscr{M})$ 
which are made of the locally decomposable multivector fields associated with the same distribution.
If ${\bf X},{\bf X}'\in\{ {\bf X}\}$ then, for every $U\subset \mathscr{M}$,
there exists a non-vanishing function $f\in\Cinfty(U)$ such that 
${\bf X}'=f{\bf X}$ on $U$.

If $\Omega\in\df^p(\mathscr{M})$ and $\mathbf{X}\in\mathfrak{X}^m(\mathscr{M})$,
the {\sl \textbf{contraction}} between ${\bf X}$ and $\Omega$ is
the natural contraction between tensor fields; in particular,
it gives zero when $p<m$ and, if $p\geq m$,
$$
 \inn({\bf X})\Omega\mid_{U}:= \sum_{1\leq i_1<\ldots <i_m\leq
 r}f^{i_1\ldots i_m} \inn(X_1\wedge\ldots\wedge X_m)\Omega 
=
 \sum_{1\leq i_1<\ldots <i_m\leq r}f^{i_1\ldots i_m} \inn
 (X_m)\ldots\inn (X_1)\Omega \ .
$$
The {\sl \textbf{Lie derivative}} of $\Omega$ with respect to ${\bf X}$ is the graded bracket (of degree $m-1$)
 $$
\Lie({\bf X})\Omega:=[\d , \inn ({\bf X})]\Omega=
(\d\inn ({\bf X})-(-1)^m\inn ({\bf X})\d)\Omega \ .
 $$
If ${\bf X}\in\vf^i(\mathscr{M})$ and ${\bf Y}\in\vf^j(\mathscr{M})$,
the {\sl \textbf{Schouten-Nijenhuis bracket}} of ${\bf X},{\bf Y}$
is the bilinear map ${\bf X},{\bf Y}\mapsto [{\bf X},{\bf Y}]$, 
where $[{\bf X},{\bf Y}]$ is a $(i+j-1)$-multivector field obtained 
as the graded commutator of $\Lie ({\bf X})$ and $\Lie ({\bf Y})$; that is,
 $$
\Lie([{\bf X},{\bf Y}])\Omega:=[\Lie ({\bf X}), \Lie ({\bf Y})]\,\Omega\ .
 $$
 This is an operation of degree $i+j-2$ which
is also called the {\sl \textbf{Lie derivative}} of ${\bf Y}$ with respect to ${\bf X}$,
and is denoted $\Lie({\bf X}){\bf Y}:=[{\bf X},{\bf Y}]$.
Furthermore, if $Y\in\vf(\mathscr{M})$ and ${\bf X}\in\vf^m(\mathscr{M})$, then
\beq
\label{liebrac}
\inn([Y,{\bf X}])\Omega=\Lie(Y)\inn({\bf X})\Omega-\inn({\bf X})\Lie(Y)\Omega \ .
\eeq

Now, let $\varrho\colon\mathscr{M}\to M$ a fiber bundle.
A multivector field $\mathbf{X}\in\mathfrak{X}^m(\mathscr{M})$ 
is \emph{$\varrho$-transverse} if,
for every $\beta\in\Omega^m(M)$ such that
$\beta_{\varrho({\rm p})}\not= 0$,
at every point ${\rm p}\in \mathscr{M}$,
we have that
$(\inn(\mathbf{X})(\bar\pi^{k*}\beta))_{{\rm p}}\not= 0$.
We are interested in the {\sl \textbf{integrable multivector fields}},
which are those locally decomposable multivector fields whose associated distribution is integrable.
Then, if $\mathbf{X}\in\mathfrak{X}^m(\mathscr{M})$ is
integrable and $\varrho$-transverse, 
its integral manifolds are local sections of the projection
$\varrho\colon \mathscr{M}\to M$.

In the particular case where $\mathscr{M}=J^1\pi$ and we have the jet bundle
$\bar\pi^1\colon J^1\pi\to M$;
an integrable multivector field $\mathbf{X}\in\vf^m (\mathscr{M})$ is said to be {\sl \textbf{holonomic}} 
if its integral sections are holonomic sections of $\bar\pi^1$.

\section*{Acknowledgments}

We want to thank Joaquim Gomis and Jordi Gaset for helpful and clarifying discussions.
We are indebted to Victor Tapia for having drawn our attention to an error in the first calculation of equation \eqref{pbraneHessian}.
We also thank the referees for their constructive comments.
We acknowledge the financial support of the 
{\sl Ministerio de Ciencia, Innovaci\'on y Universidades} (Spain), project PID2021-125515NB-C21, and the financial support for research groups AGRUPS-2022 of the Universitat Polit\`ecnica de Catalunya (UPC).


\end{document}